\documentclass[12pt]{article}

\usepackage{cite}
\usepackage{times}
\usepackage{graphicx}
\usepackage{multirow}

\title{Lattice QCD at the physical point:\\ Simulation and analysis details}

\author{
S.~D\"urr$^{1,2}$, Z.~Fodor$^{1,2,3}$, C.~Hoelbling$^{1}$, S.D.~Katz$^{1,3}$,
S.~Krieg$^{1,2}$,\\
T.~Kurth$^{1}$, L.~Lellouch$^{4}$, T.~Lippert$^{1,2}$, K.K.~Szab\'o$^{1}$,
G.~Vulvert$^{4}$
\\
\\
\normalsize{$^1$Bergische Universit\"at Wuppertal, Gaussstr.\,20, D-42119
Wuppertal, Germany.}\\[-1mm]
\normalsize{$^2$J\"ulich Supercomputing Centre, Forschungszentrum J\"ulich,
D-52425 J\"ulich, Germany.}\\[-1mm]
\normalsize{$^3$Institute for Theoretical Physics, E\"otv\"os University,
H-1117 Budapest, Hungary.}\\[-1mm]
\normalsize{$^4$Centre de Physique Th\'eorique\footnote{CPT is research unit
UMR 6207 of the CNRS and of the universities Aix-Marseille I, Aix-Marseille II
and Sud Toulon-Var, and is affiliated with the FRUMAM.}, Case 907,
Campus de Luminy, F-13288 Marseille, France.}
\\
\\
{[\,Budapest-Marseille-Wuppertal Collaboration\,]}
}

\date{}

\def\lsim{\raise0.3ex\hbox{$<$\kern-0.75em\raise-1.1ex\hbox{$\sim$}}}
\def\gsim{\raise0.3ex\hbox{$>$\kern-0.75em\raise-1.1ex\hbox{$\sim$}}}

\long\def\begincomment#1\endcomment{}

\newcommand{\bdm}{\begin{displaymath}}
\newcommand{\edm}{\end{displaymath}}
\newcommand{\beq}{\begin{equation}}
\newcommand{\eeq}{\end{equation}}
\newcommand{\bea}{\begin{eqnarray}}
\newcommand{\eea}{\end{eqnarray}}

\newcommand{\mr}{\mathrm}
\newcommand{\fm}{\,\mr{fm}}
\newcommand{\MeV}{\,\mr{MeV}}
\newcommand{\GeV}{\,\mr{GeV}}
\newcommand{\MSbar}{\overline\mr{MS}}

\newcommand{\Mpi}{M_\pi}
\newcommand{\Fpi}{F_\pi}
\newcommand{\Mka}{M_K}

\newcommand{\Met}{M_\eta}

\newcommand{\<}{\langle}
\renewcommand{\>}{\rangle}
\newcommand{\til}{\tilde}
\newcommand{\ovr}{\over}
\renewcommand{\dag}{^\dagger}

\newcommand{\be}{\beta}
\newcommand{\si}{\sigma}
\newcommand{\de}{\delta}

\begin{document}
\maketitle

\bigskip

\hspace{2mm}
\begin{minipage}{14cm}
\small
\noindent
We give details of our precise determination of the light quark masses
$m_{ud}=(m_u\!+\!m_d)/2$ and $m_s$ in $2\!+\!1$ flavor QCD, with simulated
pion masses down to 120\,MeV, at five lattice spacings, and in large volumes.
The details concern the action and algorithm employed, the HMC force with
HEX smeared clover fermions, the choice of the scale setting procedure and of
the input masses. After an overview of the simulation parameters, extensive
checks of algorithmic stability, autocorrelation and (practical) ergodicity
are reported. To corroborate the good scaling properties of our action,
explicit tests of the scaling of hadron masses in $N_f\!=\!3$ QCD are carried
out. Details of how we control finite volume effects through dedicated finite
volume scaling runs are reported. To check consistency with SU(2) Chiral
Perturbation Theory the behavior of $M_\pi^2/m_{ud}$ and $F_\pi$ as a function
of $m_{ud}$ is investigated. Details of how we use the RI/MOM procedure with a
separate continuum limit of the running of the scalar density $R_S(\mu,\mu')$
are given. This procedure is shown to reproduce the known value of $r_0m_s$ in
quenched QCD. Input from dispersion theory is used to split our value of
$m_{ud}$ into separate values of $m_u$ and $m_d$. Finally, our procedure to
quantify both systematic and statistical uncertainties is discussed.
\end{minipage}

\clearpage

\tableofcontents

\clearpage


\section{Introduction}


The goal of this paper is to give technical details of our calculation of the
average light quark mass $m_{ud}\!=\!(m_u\!+\!m_d)/2$ and of the strange quark
mass $m_s$ at the physical mass point and in the continuum \cite{Durr:2010??}.
This calculation is from first principles and sets new standards in terms of
controlling all systematic aspects of a direct calculation of quark masses.
Because the values $m_u$ and $m_d$ are also of fundamental importance, we
determine them by combining our results for $m_{ud}$ and $m_s$ with dispersive
information based on $\eta\to3\pi$ decays.
A summary of recent determinations of light quark masses in $N_f\!=\!2$ and
$N_f\!=\!2\!+\!1$ QCD is found in  \cite{Durr:2010??}.

\smallskip

Let us begin by stating the minimum requirements for a first-principles
determination of $m_{ud}$ and $m_s$ with fully controlled systematics:
\begin{enumerate}
\itemsep-2pt
\item
The action should belong to the universality class of QCD according to standard
arguments, based on locality and unitarity, and an exact algorithm should be
used.
\item
The light quark masses should be sufficiently close to their physical values
such that an extrapolation, if necessary, can be performed without adding
non-trivial assumptions.
Our simulations are performed ``\emph{at the physical mass point}'', i.e.\ with
values of $\Mpi$ and $\Mka$ which bracket the physical values; this eliminates
the need for a ``chiral extrapolation''.
\item
Simulations should be performed in volumes large enough to ensure that
finite-volume effects are well under control (we use box sizes up to
$L\!\simeq\!6\fm$).
\item
Simulations should be performed at no less than three lattice spacings $a$ to
make sure that a controlled extrapolation of all results to the continuum,
$a\to0$, can be performed.
\item
All renormalizations should be performed nonperturbatively, and the final
result should be given in a scheme which is well-defined beyond perturbation
theory (we will give our results in the RI/MOM scheme).
\item
The scale and other input masses should be set by quantities whose relation to
experiment are direct and transparent (we use the masses of the $\Omega$
baryon, the pion and the kaon).
\end{enumerate}

\smallskip

The present work contains additional innovative features which are not required
to give an ab-initio result, but help to keep all systematic errors small:
\begin{enumerate}
\itemsep-2pt
\setcounter{enumi}{6}
\item
We devise a method, tailored to needs of studies with Wilson-like fermions, to
reconstruct the renormalized quark masses $m_{ud}$ and $m_s$ from the much
simpler renormalization of the quantities $m_s/m_{ud}$ and $m_s\!-\!m_{ud}$.
We call it the ``ratio-difference method''.
\item
We propose an approach which overcomes the RI/MOM ``window-problem''.
It is based on taking a separate continuum limit of the evolution function
$R_S(\mu,\mu')$ of the scalar density $S$ from a scale $\mu\!\sim\!2\GeV$,
where the RI/MOM procedure yields reliable results, to a scale
$\mu'\!\sim\!4\GeV$ where one may make (controlled) contact with perturbation
theory.
\item
We use an advanced analysis procedure to assess the size of both the
statistical and the systematic uncertainties (the same one as in
\cite{Durr:2008zz}).
\end{enumerate}
In our view, item 2 marks the beginning of a new era in numerical lattice QCD,
because it avoids an extrapolation in quark masses which, generically, requires
strong assumptions, thus relinquishing the first-principles approach (see the
discussion in \cite{Durr:2010??}).

\begin{figure}[tb]
\centering
\includegraphics[width=12.0cm]{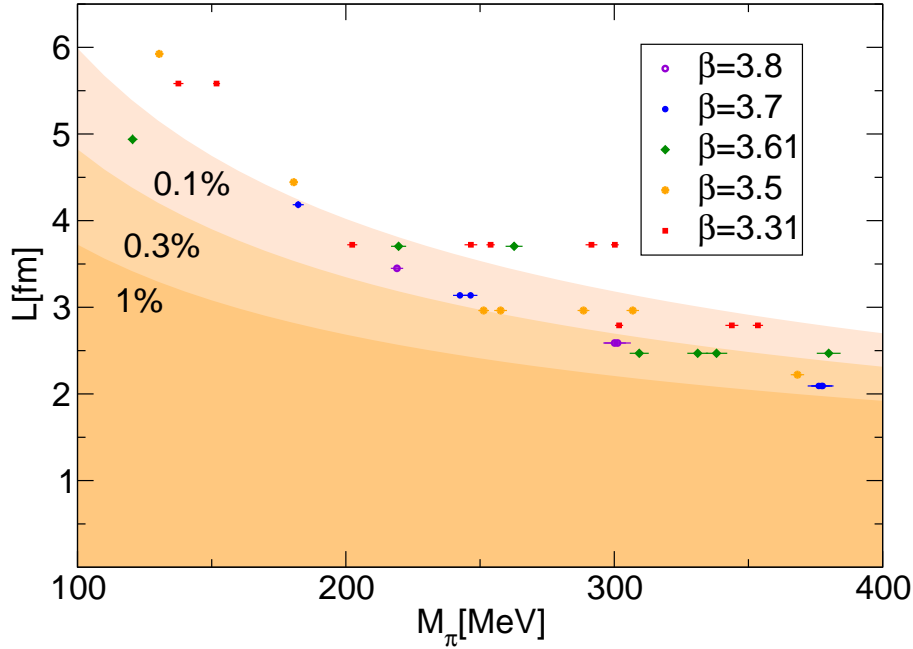}
\caption{\label{fig:landscape}\sl
Summary of our simulation points. The pion masses and the spatial sizes of the
lattices are shown for our five lattice spacings. The percentage labels
indicate regions, in which the expected finite volume effect
\cite{Colangelo:2005gd} on $\Mpi$ is larger than 1\%, 0.3\% and 0.1\%,
respectively. In our runs this effect is smaller than about 0.5\%, but we
still correct for this tiny effect.}
\end{figure}

To give the reader an overview of where we are in terms of simulated pion
masses $\Mpi$ and spatial box sizes $L$, a graphical survey of (some of) our
simulation points is provided in Fig.\,\ref{fig:landscape} (with more details
given in Sec.\,5).
We have data at 5 lattice spacings in the range $0.054\!-\!0.116\fm$,
with pion masses down to $\sim\!120\MeV$ and box sizes up to $\sim\!6\fm$.
Comparison with Chiral Perturbation suggests that our finite volume effects
are typically below 0.5\%, and close to the physical mass point (which is
the most relevant part) even smaller.
Still, we correct for them by means of Chiral Perturbation Theory
\cite{Colangelo:2005gd}, and test the correctness of this prediction through
explicit finite volume scaling runs (see below).

The remainder of this paper is organized as follows.
In Sec.\,2 details are given concerning the action and algorithm employed,
while Sec.\,3 specifies how one determines the HMC force with HEX smeared
clover fermions.
Our choice of the scale setting procedure and of the input masses is discussed
in Sec.\,4, with simulation parameters tabulated in Sec.\,5.
Checks of algorithmic stability are summarized in Sec.\,6, while
autocorrelation and (practical) ergodicity issues are reported in Sec.\,7.
To corroborate the good scaling properties of our action, explicit tests of the
scaling of hadron masses in $N_f\!=\!3$ QCD are carried out, see Sec.\,8.
Details of how we control finite volume effects through dedicated finite volume
scaling runs are reported in Sec.\,9.
To test consistency with SU(2) Chiral Perturbation Theory the behavior of
$M_\pi^2/m_{ud}$ and $F_\pi$ as a function of $m_{ud}$ is investigated in
Sec.\,10.
Details of how we use the RI/MOM procedure with a separate continuum limit of
the running of the scalar density $R_S(\mu,\mu')$ are given in Sec.\,11.
To show the reliability of this procedure the known value of $r_0m_s$ in
quenched QCD is reproduced, see Sec\,12.
In Sec.\,13 it is discussed how one may use input from dispersion theory to
split our value of $m_{ud}$ into separate values of $m_u$ and $m_d$.
Sec.\,14 specifies our procedure to quantify both systematic and statistical
uncertainties.
Our final result is summarized in Sec.\,15.


\section{Action and algorithm details}


We use a tree-level $O(a^2)$-improved Symanzik gauge action
\cite{Luscher:1985zq} together with tree-level clover-improved Wilson
fermions \cite{Sheikholeslami:1985ij}, coupled to links which have undergone
two levels of HEX smearing.
The latter derives from the HYP setup \cite{Hasenfratz:2001hp}, but with
stout/EXP smearing \cite{Morningstar:2003gk} as the effective ingredient~--
see \cite{Capitani:2006ni} for details.
In terms of the original [$U_\mu(x)$] and smeared [$V_\mu(x)$] gauge links (see
below) our action takes the form \cite{Luscher:1985zq,Sheikholeslami:1985ij}
\bea
S&=&S_g^\mr{Sym}+S_f^\mr{SW}
\nonumber
\\
S_g^\mr{Sym}&=&\be\;
\Big[
\frac{c_0}{3}\,\sum_\mr{plaq}\mr{Re\,Tr\,}(1-U_\mr{plaq})+
\frac{c_1}{3}\,\sum_\mr{rect}\mr{Re\,Tr\,}(1-U_\mr{rect})
\Big]
\nonumber
\\
S_f^\mr{SW}&=&S_f^\mr{W}[V]-\frac{c_\mr{SW}}{2}\,
\sum_x\,\sum_{\mu<\nu}\,(\overline{\psi}\,\si_{\mu\nu}F_{\mu\nu}[V]\,\psi)(x)
\label{def_action}
\eea
with $\si_{\mu\nu}\!=\!\frac{\mr{i}}{2}[\gamma_\mu,\gamma_\nu]$ and
$S_f^\mr{W}$ denoting the standard Wilson action, and where the expression for
the field strength can be found in \cite{Sheikholeslami:1985ij}.
In $S_g^\mr{Sym}$ only the original thin links $U_\mu(x)$ are used.
The parameters $c_0,c_1$ \cite{Luscher:1985zq} and $c_\mr{SW}$
\cite{Sheikholeslami:1985ij} are set to their tree-level values
\beq
c_1=-1/12\;,\quad
c_0=1-8c_1=5/3\;,\quad
c_\mr{SW}=1\;.
\eeq
Note that both the hopping part and the clover term in $S_f^\mr{SW}$ use the
same type of HEX-smeared links $V_\mu(x)\equiv V^{(2)}_\mu(x)$.
Those are constructed from the thin links $V^{(0)}_\mu(x)\equiv U_\mu(x)$ via
\cite{Capitani:2006ni}
\bea
\Gamma_{\mu,\nu\rho}^{(1,n)}(x)&=&
\sum_{\pm\si\neq(\mu,\nu,\rho)}
V^{(n-1)}_{\si}(x)
V^{(n-1)}_{\mu}(x\!+\!\hat\si)
V^{(n-1)}_{\si}(x\!+\!\hat\mu)\dag
\nonumber
\\[2pt]
V_{\mu,\nu\rho}^{(1,n)} (x)&=&
\exp\Big(\frac{\alpha_3}{2}\,
P_\mr{TA}^{}\Big\{
\Gamma_{\mu,\nu\rho}^{(1,n)}(x) V^{(n-1)}_{\mu}(x)\dag
\Big\}
\Big)V_\mu^{(n-1)}(x)
\nonumber
\\[4pt]
\Gamma_{\mu,\nu}^{(2,n)}(x)&=&
\sum_{\pm\si\neq(\mu,\nu)}
V_{\si,\mu\nu}^{(1,n)}(x)
V_{\mu,\nu\si}^{(1,n)}(x\!+\!\hat\si)
V_{\si,\mu\nu}^{(1,n)}(x\!+\!\hat\mu)\dag
\nonumber
\\[2pt]
V_{\mu,\nu}^{(2,n)} (x)&=&
\exp\Big(\frac{\alpha_2}{4}\,
P_\mr{TA}^{}\Big\{
\Gamma_{\mu,\nu}^{(2,n)}(x) V^{(n-1)}_{\mu}(x)\dag
\Big\}
\Big)V_\mu^{(n-1)}(x)
\nonumber
\\[4pt]
\Gamma_{\mu}^{(3,n)}(x)&=&
\sum_{\pm\nu\neq(\mu)}
V_{\nu,\mu}^{(2,n)}(x)
V_{\mu,\nu}^{(2,n)}(x\!+\!\hat\nu)
V_{\nu,\mu}^{(2,n)}(x\!+\!\hat\mu)\dag
\nonumber
\\[2pt]
V_\mu^{(3,n)}(x)&=&
\exp\Big(\frac{\alpha_1}{6}\,
P_\mr{TA}^{}\Big\{
\Gamma_{\mu}^{(3,n)}(x) V^{(n-1)}_{\mu}(x)\dag
\Big\}
\Big)V_\mu^{(n-1)}(x)
\;\equiv\;V_\mu^{(n)}(x)
\label{def-hex}
\eea
without summation over repeated indices.
Here
\beq
P_\mr{TA}\{M\}=\frac{1}{2}[M\!-\!M\dag]-\frac{1}{6}\mr{Tr}[M\!-\!M\dag]
\label{eq:hex_ta}
\eeq
denotes the traceless anti-hermitean part of the $3\!\times\!3$ matrix $M$.
We use the parameters $\alpha_1=0.95$, $\alpha_2=0.76$ and $\alpha_3=0.38$.
In terms of the standard stout/EXP smearing convention
\cite{Morningstar:2003gk}
\bea
\Gamma_\mu^{(n)}(x)&=&\sum_{\pm\nu\neq\mu}
V_\nu^{(n-1)}(x)V_\mu^{(n-1)}(x\!+\!\hat\nu)V_\nu^{(n-1)}(x\!+\!\hat\mu)\dag
\nonumber\\
V_\mu^{(n)}(x)&=&
\exp\Big(\rho\,
P_\mr{TA}^{}
\Big\{
\Gamma_\mu^{(n)}(x)V_\mu^{(n-1)}(x)\dag
\Big\}
\Big)V_\mu^{(n-1)}(x)
\label{def-stout}
\eea
the values of $\alpha_i$, $i=1,2,3$, above correspond to
$\rho$=0.158333, 0.19 and 0.19, respectively.
This smearing differs from the one we used in \cite{Durr:2008rw,Durr:2008zz}
in that the fermions interact even more locally with the gauge fields
here (cf.\ the discussion in the supplementary online material of
\cite{Durr:2008zz} and the appendix of \cite{Capitani:2006ni}).
Note, finally, that our action is exactly as ultralocal (in position space) as
the original Wilson/clover action, since $D(x,y)\!=\!0$ for $|x\!-\!y|\!>\!1$.

As we use the hybrid Monte-Carlo (HMC) algorithm \cite{Duane:1987de}, a
non-trivial ingredient with this action is the coding of the molecular
dynamics (MD) force, which will be discussed in the next section.
The MD updates are performed in quadruple precision, to ensure exact
reversibility in our target (double precision) accuracy.
Further particulars of our implementation~--
even-odd preconditioning \cite{DeGrand:1990dk}, multiple
time-scale integration (``Sexton-Weingarten scheme'') \cite{Sexton:1992nu},
mass preconditioning (``Hasenbusch trick'') \cite{Hasenbusch:2001ne},
Omelyan integrator \cite{Takaishi:2005tz},
RHMC acceleration with multiple pseudofermions \cite{Clark:2006fx} and
mixed-precision solver \cite{Durr:2008rw}~--
have been described in \cite{Durr:2008rw}.
As has been noted in the literature \cite{Urbach:2005ji,Aoki:2008sm},
combining several of these ingredients yields a dramatic reduction of the
critical slowing down that has traditionally been observed for light quark
masses.
As we show in this paper, the thorough combination of all these ingredients
allows for simulations directly at the physical mass point, in large volumes,
with several lattice spacings, such that a controlled extrapolation to the
continuum can be performed.


\section{HMC force with smeared links}


\newcommand{\der}[2]{\ensuremath{\frac{\partial #1}{\partial #2}}}
\newcommand{\derd}[2]{\ensuremath{\frac{\delta #1}{\delta #2}}}
\newcommand{\derdir}[3]{\left[\frac{\delta #1}{\delta #2}\right]_{#3}}
\newcommand{\tr}{\mathrm{Tr}}
\newcommand{\Vn}{V^{(0)}}
\newcommand{\Vnd}{V^{(0)\dagger}}
\newcommand{\Vi}{V^{(1)}}
\newcommand{\Vid}{V^{(1)\dagger}}
\newcommand{\Vii}{V^{(2)}}
\newcommand{\Viid}{V^{(2)\dagger}}
\newcommand{\Viii}{V^{(3)}}
\newcommand{\Cn}{C^{(1)}}
\newcommand{\Ci}{C^{(2)}}
\newcommand{\Cii}{C^{(3)}}
\newcommand{\Sn}{\Sigma^{(0)}}
\newcommand{\Si}{\Sigma^{(1)}}
\newcommand{\Sii}{\Sigma^{(2)}}
\newcommand{\Siii}{\Sigma^{(3)}}
\newcommand{\taproj}{P_\mathrm{TA}^{}} 

We use two steps of HEX smearing \cite{Capitani:2006ni} in our fermion action
$S_f$, both for the Wilson and for the clover term.
Our $S_f$ depends on the thin (unsmeared) links only through the smeared links
\beq
S_f=S_f(V^{(2)}(V^{(1)}(V^{(0)}\!=\!U)))
\label{Sf_dependence}
\eeq
where $V^{(n)}$ denotes the HEX smeared links, with $n$ indicating the smearing
level.
Generically, the fermionic contribution to the HMC force is given by the
gauge derivative $\de{S_f}/\de{U}$.
In order to obtain the derivative of $S_f$ with respect to $U$ for our two-step
smearing recipe, we will apply the chain rule twice, which leads us to the
following scheme.
First one calculates
\beq
\derd{S_f}{V^{(2)}}
\eeq
which encodes the details of how the fermions are coupled to the smeared gauge
fields.
This part of the calculation is not related to the smearing, one just takes
$\de{S_f[U]}/\de{U}$ and replaces $U\to V^{(2)}$.
The main consequence of the nested dependence (\ref{Sf_dependence}) is the
recursion formula
\beq
\derd{S_f}{V^{(n-1)}}=\derd{S_f}{V^{(n)}}\star\derd{V^{(n)}}{V^{(n-1)}}
\label{eq:hex_sch}
\eeq
where the proper definition of the star-product and of the second term will be
given below.
Thinking in terms of routines of the computer code, one such step takes the
previous derivative $\de{S_f}/\de{V^{(n)}}$ and the links $V^{(n)},V^{(n-1)}$
as input and yields the next derivative $\de{S_f}/\de{V^{(n-1)}}$.
This procedure needs to be called $n$ times, at the end we obtain the final
fermion force
\beq
\derd{S_f}{U}=\derd{S_f}{V^{(0)}}
\;.
\eeq
Since the extension to a second and possibly more steps is straightforward,
we will only consider the derivation for one level of HEX smearing in the
following.

We now specify the two main ingredients in the derivation of the HMC force for
fat-link actions, that is the gauge derivative and the pertinent chain rule.
Since an $SU(3)$ matrix in the fundamental representation is a complex
$3\times3$ matrix with only 8 independent degrees of freedom, it is a
structured matrix and the derivative has to be defined properly.

\emph{The Lie algebra.}
For $U\!\in\!SU(3)$ an infinitesimal change can be written as
\beq
U \to U^\prime=\exp\Big(\sum_A u_A T_A\Big)U
\label{eq:hex_up}
\eeq
with real parameters $u_A$ and the set $\{T_A|A=1\dots8\}$ forming a basis in
the space of the traceless, anti-hermitian matrices, i.e.\ of the tangent space
of the group.
These matrices are normalized through $\tr(T_AT_B)=-\delta_{AB}$.
Using the trace, one can define a scalar product on this vector space; for
$X=\sum_A x_A T_A$ and $Y=\sum_A y_A T_A$ in the Lie algebra the scalar
product
\beq
\langle X,Y \rangle=-\tr(X Y)=\sum_A x_A y_A
\label{eq:hex_sp}
\eeq
allows one to build a projector which restricts an arbitrary $3\!\times\!3$
matrix, $M$, to the tangent space
$\taproj(M)=-\sum_A T_A \mr{Re\,Tr}(T_A M)$, with $\taproj(M)$ defined in
(\ref{eq:hex_ta}).
Furthermore, it is easy to show that for arbitrary matrices $M$ and $N$
\beq
\mr{Re\,Tr}(\taproj(M)N)=\mr{Re\,Tr}(\taproj(N)M)
\;.
\label{eq:hex_mn}
\eeq

\emph{Complex valued functions.}
Let $f$ be a complex valued function on the group $SU(3)$.
Its derivative with respect to the group element is a vector in the Lie algebra
\beq
\derd{f}{U}= \sum_A T_A \derdir{f}{U}{A}
\label{eq:hex_def}
\eeq
where the components are defined as
\beq
\derdir{f}{U}{A}=
\der{f(U^\prime)}{u_A}=
\lim_{u_A\to 0} \Big[ f(\exp(u_A T_A)U) - f(U) \Big]/u_A=
\tr \Big(T_A U \der{f}{U^T} \Big)
\;.
\eeq
Throughout, the partial derivative with respect to $U$ under the trace is to
be understood as a derivative with respect to unconstrained matrix elements.
In particular, this means that
\beq
\der{U_{cd}}{U_{ab}}= \delta_{ca}\delta_{bd}
\;.
\eeq
If $f$ depends on the adjoint matrix $U\dag$, then using the identity
$U\dag=U^{-1}$ this dependence is converted into a dependence on $U$, with the
consequence that
\beq
\der{U\dag_{cd}}{U_{ab}}= -U\dag_{ca}U\dag_{bd}
\;.
\eeq

\emph{Real valued functions.}
For a real valued function $f$ the group derivative takes the form
\beq
\derd{f}{U}= \sum_A T_A \derdir{f}{U}{A}=
\sum_A T_A \tr \Big(T_A U \der{f}{U^T} \Big) =
-\taproj \Big(U \der{f}{U^T}\Big)
\;.
\eeq

\emph{SU(3) valued functions.}
Let $V\!\in\!SU(3)$ be a function of $U$, such that $V:SU(3)\to SU(3)$ is a
mapping on the group space.
If $U$ changes as in (\ref{eq:hex_up}), with small parameters $u_A=O(\epsilon)$,
then $V$ will also undergo a small change, which may be written as
\beq
V\to V^\prime= V(U^\prime)= \exp(\sum_B v_B T_B)V
\eeq
where the small parameters $v_B=O(\epsilon)$ are real-valued functions of the
original parameters $u_A$.
Below we shall need the $u_A$ derivative of $V^\prime$, that is
\beq
\der{V^\prime}{u_A}=\der{(\sum_B v_B(u) T_B)}{u_A}V
\eeq
which, upon taking the limit $u_A\to0$, implies
\beq
\derdir{V}{U}{A}=\sum_B \der{v_B}{u_A}\bigg|_{u=0}T_B V
\;.
\label{eq:hex_vua}
\eeq

\emph{Chain rule.}
Let the function $f$ depend on $U$ only via $V$, and let us calculate its
derivative with respect to $U$.
Again, $U$ transforms with infinitesimal parameters $u_A$, resulting in an
infinitesimal change of the variables $v_B=v_B(u)$ of $V$.
The usual chain rule yields
\beq
\der{f}{u_A}=\sum_B \der{f}{v_B}\der{v_B}{u_A}
\eeq
and, after taking the limit $u_A\to0$, we arrive at
\beq
\derdir{f}{U}{A}=\sum_B \derdir{f}{V}{B}\der{v_B}{u_A}\bigg|_{u=0}
\;.
\eeq
With (\ref{eq:hex_sp}) and the definition of the gauge derivative from
(\ref{eq:hex_def}, \ref{eq:hex_vua}), this may be rewritten as
\beq
\derdir{f}{U}{A}=- \tr\Big(\derd{f}{V}\derdir{V}{U}{A} V\dag\Big)
\;.
\label{eq:hex_fua}
\eeq
This formula is the chain rule for the gauge derivative, which can formally
be stated as
\beq
\derd{f}{U}=\derd{f}{V} \star \derd{V}{U}
\;.
\eeq

With these ingredients we can now specify the HMC force for a fermion action
with one step of HEX smearing.
In the following we will simplify our notation by replacing $V^{(i,n)}$ with
$V^{(i)}$.
One HEX smearing, $U_\mu \to \Viii_{\mu}$, is built from three substeps (cf.\
Eq.\,\ref{def-hex})
\bea
\Vi_{\mu,\nu,\rho}(x) &=&
\exp\Big(\taproj[\Cn_{\mu,\nu,\rho}(x) U\dag_\mu(x)]\Big)
\cdot U_\mu(x)
\nonumber
\\
\Vii_{\mu,\nu}(x) &=&
\exp\Big(\taproj[\Ci_{\mu,\nu}(x) U\dag_\mu(x)]\Big)
\cdot U_\mu(x)
\nonumber
\\
\Viii_{\mu}(x) &=&
\exp\Big(\taproj[\Cii_{\mu}(x) U\dag_\mu(x)]\Big)
\cdot U_\mu(x)
\eea
where $\taproj$ has been defined in (\ref{eq:hex_ta}) and $\Cn$, $\Ci$, $\Cii$
represent staples constructed via
\bea
\Cn_{\mu,\nu,\rho}(x) &=&
\frac{\alpha_3}{2}\sum_{\pm\si\ne(\mu,\nu,\rho)}
U_\si(x) U_\mu(x\!+\!\hat\si) U\dag_\si(x\!+\!\hat\mu)
\nonumber
\\
\Ci_{\mu,\nu}(x) &=&
\frac{\alpha_2}{4}\sum_{\pm\si\ne(\mu,\nu)} \Vi_{\si,\mu,\nu}(x)
\Vi_{\mu,\si,\nu}(x\!+\!\hat\si) \Vid_{\si,\mu,\nu}(x\!+\!\hat\mu)
\nonumber
\\
\Cii_{\mu}(x) &=&
\frac{\alpha_1}{6}\sum_{\pm\si\ne\mu} \Vii_{\si,\mu}(x)
\Vii_{\mu,\si}(x\!+\!\hat\si) \Viid_{\si,\mu}(x\!+\!\hat\mu)
\eea
with the factors $\alpha_i/(2(d\!-\!1))$ included (for reasons to become
obvious below).
In the following we will drop Lorentz indices and space-time arguments for
simplicity.

The task is to calculate the HMC force [with $V=\Viii$]
\beq
\derd{S_f}{U}= \derd{S_f}{V} \star \derd{V}{U},
\eeq
where $\de{S_f}/\de{V}$ is already known.
The $A$ component of the star product reads [see (\ref{eq:hex_fua})]
\beq
\derdir{S_f}{U}{A}=
-\Big[V^{(3)\dagger} \derd{S_f}{\Viii}\Big]_{ab} \derdir{\Viii_{ba}}{U}{A}=
-\Siii_{ab} \tr \Big(T_A U \der{\Viii_{ba}}{U^T}\Big)
\eeq
where $\Siii$ contains the part of the force which is already known
\beq
\Siii= V^{(3)\dagger}\derd{S_f(\Viii)}{\Viii}
\;.
\eeq
Since $S_f$ is a real-valued function we can write this in the compact form
\beq
\derd{S_f}{U}= \taproj \Big( U \Siii_{ab} \der{\Viii_{ba}}{U^T}\Big)
\;.
\eeq

To improve readability, let us temporarily denote $V^{(3)}$ by $V$.
The last substep is then $V=\exp(A)U$, where $A=\taproj(CU\dag)$ and thus
\beq
\derd{S_f}{U}=
\taproj\Big(U\Sigma_{ab} \der{V_{ba}}{U^T}\Big)=
\taproj\Big(U\Sigma\exp(A)\Big) +
\taproj\Big(U\Sigma_{ab} \der{\exp(A)_{bc}}{U^T}U_{ca}\Big)
\;.
\eeq
The derivative of the exponential of a traceless anti-hermitian $A$ reads
[see Eq.\,(68) of \cite{Morningstar:2003gk}]
\beq
d(\exp(A))=
\tr(A\cdot dA)B_1+\tr(A^2\cdot dA)B_2+f_1 dA+f_2A\cdot dA+f_2 dA\cdot A
\eeq
with $B_{1,2}$ being second-order polynomials of $A$ and $f_{1,2}$ complex
constants which depend on the trace and determinant of $A$.
Using the cyclicity of the trace in the color indices, we arrive at
\begin{eqnarray}
&&
\taproj\Big(U\Sigma_{ab}\der{V_{ba}}{U^T}\Big)=
\taproj\Big(U\Sigma\exp(A)\Big)+
\nonumber
\\
&&
\taproj\Big(U\der{A_{ab}}{U^T}\cdot[
\mr{Tr}(U\Sigma B_1)A +
\mr{Tr}(U\Sigma B_2)A^2 + f_1 U\Sigma + f_2 U\Sigma A + f_2 AU\Sigma]_{ba}\Big)
\label{eq:hexfor_sqbra}
\end{eqnarray}
where the second term contains the derivative of $A\!=\!\taproj(CU\dag)$.
We use the identity
\beq
\taproj\Big(U\der{\taproj(M)_{ab}}{U^T}N_{ba}\Big) =
\taproj\Big(U\der{M_{ab}}{U^T}\taproj(N)_{ba}\Big)
\eeq
valid for arbitrary $M$ and $N$ [see (\ref{eq:hex_mn})] to shuffle the
projector in $A\!=\!\taproj(CU\dag)$ to the matrix in square brackets in
(\ref{eq:hexfor_sqbra}).
Next, we use that the derivative of $CU\dag$ can be written as
\beq
U\der{(CU\dag)_{ab}}{U^T}\taproj[...]_{ba}=
U\der{C_{ab}}{U^T}(U\dag\taproj[...])_{ba} -
\taproj[...]\cdot CU\dag
\eeq
and introduce a convenient notation for the expression in square brackets
in (\ref{eq:hexfor_sqbra}) by means of
\beq
Z =U\dag \taproj\Big[
\tr(U\Sigma B_1)A +
\tr(U\Sigma B_2)A^2 + f_1 U\Sigma + f_2 U\Sigma A + f_2 AU\Sigma \Big]
\;.
\eeq
With this at hand, and using the relation $\exp(A)\!=\!VU\dag$, we arrive at
the compact expression
\beq
\taproj\Big(U\Sigma_{ab}\der{V_{ba}}{U^T}\Big)=
\taproj\Big(U(\Sigma V\!-\!ZC)U\dag+U\der{C_{ab}}{U^T}Z_{ba}\Big)
\;.
\label{compact}
\eeq

Finally, we reintroduce the superscript $(3)$ and note that the $U$
dependence of $\Cii$ comes exclusively from $\Vii$.
With these adjustments, relation (\ref{compact}) takes the form
\beq
\taproj\Big(U\Siii_{ab} \der{\Viii_{ba}}{U^T}\Big) =
\taproj\Big( U (\Siii \Viii - Z^{(3)} C^{(3)}) U\dag \Big) +
\taproj\Big(U\Sii_{ab} \der{\Vii_{ba}}{U^T}\Big)
\label{eq:32}
\eeq
where $\Sii$ is defined as
\beq
\Sii_{ab}= \der{\Cii_{cd}}{\Vii_{ba}}Z^{(3)}_{dc}
\eeq
meaning that the term $\Sigma^{(2)}$ can be calculated in the similar way
as $\Sigma^{(3)}$.
This procedure can be iterated, and we find for an action with one step of HEX
smearing
\beq
\derd{S_f}{U}=
\sum_{i=3,2,1}\taproj\Big(U(\Sigma^{(i)}V^{(i)}-Z^{(i)}C^{(i)})U\dag\Big)+
\taproj\Big(U\Sn\Big)
\eeq
where $\Sigma^{(i)}$ is defined as
\beq
\Sigma^{(i)}_{ab}=\der{C^{(i+1)}_{cd}}{V^{(i)}_{ba}}Z^{(i+1)}_{dc}
\quad\mr{for}\quad
i=0,1,2
\;.
\eeq
The object $\partial{C^{(i+1)}}/\partial{V^{(i)}}$ is a straightforward
staple derivative, where some care needs to be taken w.r.t.\ the Lorentz
indices.
With this formula, one may implement a routine which calculates the
HMC force for a fermion action with one step of HEX smearing.
The extension to a second smearing step is realized through a second call to
this routine as shown in (\ref{Sf_dependence}).

This calculation of the HMC force with HEX smeared fermion actions extends the
results of \cite{Morningstar:2003gk,Hasenfratz:2007rf}.
An early treatise of the HMC force for fat-link fermion actions is found in
\cite{Kamleh:2004xk}.


\section{Scale setting and input masses}


To set the scale and to adjust the light and strange quark masses
$m_{ud}\!=\!(m_u\!+\!m_d)/2$ and $m_s$ to their physical values, we need to
identify three quantities which can be precisely computed on the lattice and
measured in experiment.
We will use the mass of the $\Omega$ baryon and of the pseudoscalar mesons
$\pi,K$ for this purpose, in the latter case with a small correction for
isospin-breaking and electromagnetic effects (see below).

In other words, at the point where $\Mpi/M_\Omega$ and $\Mka/M_\Omega$ agree
with the physical values of these ratios, the measured value of $aM_\Omega$ is
identified with the lattice spacing times $1.672\GeV$ \cite{Nakamura:2010zzi}, and
this yields $a^{-1}$.
In \cite{Durr:2008zz} we used also the $\Xi$ baryon to set the scale.
As discussed there, correlation functions of spin $3/2$ baryons are somewhat
noisier than those of spin $1/2$ baryons.
On the other hand, the more light valence quarks present in a baryon, the
larger the fluctuations.
In \cite{Durr:2008zz} with $\Mpi$ down to $190\MeV$ these two effects balanced
each other, rendering the $\Omega$ and the $\Xi$ equally good choices.
In the present paper we go down to $\Mpi\!=\!120\MeV$, and the $\Omega$ with
no light valence quark is the better choice.
We use the standard mass-independent scale-setting scheme, in which this
lattice spacing is subsequently attributed to all ensembles with the same
coupling $\beta\!=\!6/g_0^2$ and $N_f$.

Our ensembles bracket the physical quark masses $m_{ud}$ and $m_s$ in
the sense that the span of simulated $\Mpi^2$ and
$M_{s\bar{s}}^2\!\equiv\!2\Mka^2\!-\!\Mpi^2$ encompass the physical values
given below.
As a result, it suffices to use a parametrization of $aM_\Omega$ as a function
of $(a\Mpi)^2$ and $(aM_{s\bar{s}})^2$ which describes the entire data set.
We find that the Taylor ansatz
$aM_\Omega=c_0+c_1(a\Mpi)^2+c_2(aM_{s\bar{s}})^2+c_3(aM_{s\bar{s}})^4$ works
perfectly.

\begincomment 

Since our lattice simulations are performed in the isospin symmetric limit
$m_d\!=\!m_u$ and do not account for electromagnetic interactions, the physical
input meson masses must be corrected for these effects.
A detailed account of this procedure is given in \cite{FLAG}; here we include
an abridged version of this discussion for the reader's convenience.

Let us first specify the notation.
For a given particle $P\!\in\!\{\pi^+,\pi^0,K^+,K^0\}$ we will separate its
physical mass $M_P$ into a QCD contribution $\hat{M}_P$ and a QED contribution
$M_P^\gamma$ which together make up for $M_P\!=\!\hat{M}_P\!+\!M_P^\gamma$.
While this splitting is slightly ambiguous in principle, the numerical
implications are far too small to matter from any practical viewpoint
\cite{FLAG}.
Since the self-energies of the Nambu-Goldstone bosons diverge in the chiral
limit, it is more appropriate to consider the electromagnetic contribution to
the square of the mass, i.e.\
\beq
\Delta_P^\gamma\equiv M_P^2-\hat{M}_P^2=2M_PM_P^\gamma+O(e^4)
\;.
\eeq

The key ingredient in what follows is that the main effect of electromagnetic
interactions is to lift the mass of the charged mesons, while the neutral ones
stay virtually unchanged.
This picture is formalized in ``Dashen's theorem'' \cite{Dashen:1969eg} which
states that, to leading order in the chiral expansion,
$\Delta_{\pi^0}^\gamma=\Delta_{K^0}^\gamma=0$, while 
$\Delta_{\pi^+}^\gamma=\Delta_{K^+}^\gamma>0$.
It is convenient to express both the electromagnetic splitting and the effect
of isospin violation in QCD in terms of the experimental splitting
$\Delta_\pi\equiv M_{\pi^+}^2-M_{\pi^0}^2$.
This triggers the definitions
\beq
\Delta_{\pi^0}^\gamma\equiv\epsilon_{\pi^0}\Delta_\pi\;,\qquad
\Delta_{  K^0}^\gamma\equiv\epsilon_{  K^0}\Delta_\pi\;,\qquad
\hat{M}_{\pi^+}^2-\hat{M}_{\pi^0}^2=\epsilon_m\Delta_\pi
\eeq
which then yield the compact relations
\beq
\Delta_{\pi^+}^\gamma=(1+\epsilon_{\pi^0}-\epsilon_m)\Delta_\pi\;,\qquad
\Delta_{K^+}^\gamma=(1+\epsilon+\epsilon_{K^0}-\epsilon_m)\Delta_\pi
\eeq
where we have introduced the dimensionless coefficient
\beq
\epsilon\equiv(\Delta_{K^+}^\gamma-\Delta_{K^0}^\gamma-
\Delta_{\pi^+}^\gamma+\Delta_{\pi^0}^\gamma)/\Delta_\pi
\eeq
to quantify the violations of Dashen's theorem.
To determine the numerical values of the coefficients involved is a non-trivial
task; in the following we use the values advocated by FLAG \cite{FLAG}
\beq
\epsilon=0.7(5)\;,\qquad
\epsilon_{\pi^0}=0.07(7)\;,\qquad
\epsilon_{K^0}=0.3(3)\;,\qquad
\epsilon_m=0.04(2)\;.
\label{val_epsilon}
\eeq
For the electromagnetic contributions to the light pseudoscalar meson masses
this means \cite{FLAG}
\beq
\begin{array}{ccc}
M_{\pi^+}^\gamma=4.7(3)\MeV\;,& M_{\pi^0}^\gamma=0.3(3)\MeV\;,&
M_{\pi^+}^\gamma-M_{\pi^0}^\gamma=4.4(1)\MeV\;,
\\[1mm]
M_{  K^+}^\gamma=2.5(7)\MeV\;,& M_{  K^0}^\gamma=0.4(4)\MeV\;,&
M_{  K^+}^\gamma-M_{  K^0}^\gamma=2.1(6)\MeV\;.
\end{array}
\eeq
Note that the self-energy difference between the charged and the neutral pion
depends on the same coefficient $\epsilon_m$ that describes the mass difference
in QCD~-- this is why the prediction for $M_{\pi^+}^\gamma-M_{\pi^0}^\gamma$ is
so precise.
Subtracting these contributions from the PDG values \cite{Nakamura:2010zzi} yields
\cite{FLAG}
\beq
\begin{array}{ccc}
\hat{M}_{\pi^+}=134.8(3)\MeV\;,& \hat{M}_{\pi^0}=134.6(3)\MeV\;,&
\hat{M}_{\pi^+}-\hat{M}_{\pi^0}=+0.2(1)\MeV\;,
\\[1mm]
\hat{M}_{  K^+}=491.2(7)\MeV\;,& \hat{M}_{  K^0}=497.2(4)\MeV\;,&
\hat{M}_{  K^+}-\hat{M}_{  K^0}=-6.1(6)\MeV
\end{array}
\eeq
in QCD without electromagnetic interactions.

The last step is to perform the limit $m_d\!-\!m_u\!\to\!0$ at fixed
$m_u\!+\!m_d$.
Let us denote the pion and kaon masses in that limit by $\bar{M}_\pi$ and
$\bar{M}_K$, respectively.
Since the operation $u\!\leftrightarrow\!d$ interchanges
$\pi^+\!\leftrightarrow\!\pi^-$ and $K^+\!\leftrightarrow\!K^0$, the expansion
of the quantities $\hat{M}_{\pi^+}^2$ and
$\frac{1}{2}(\hat{M}_{K^+}^2\!+\!\hat{M}_{K^0}^2)$ in $m_d\!-\!m_u$ contains
only even powers.
As shown in \cite{Gasser:1983yg} the effects on
$\hat{M}_{\pi^+}^2\!-\!\bar{M}_\pi^2$ generated by
$m_d\!-\!m_u$ are of order $O([m_d\!-\!m_u]^2[m_d\!+\!m_u])$ and thus small
compared to the difference $\hat{M}_{\pi^+}^2\!-\!\hat{M}_{\pi^0}^2$
for which an estimate has been given above.
In the case of
$\frac{1}{2}(\hat{M}_{K^+}^2\!+\!\hat{M}_{K^0}^2)\!-\!\bar{M}_K^2$ the
expansion does contain an NLO contribution whose size is determined by
$2L_8\!-\!L_5$, but knowledge of these low-energy constants shows that this
contribution is rather small, too.
As a result, the effects generated by $m_d\!-\!m_u$ in
$\hat{M}_{\pi^+}^2\!-\!\bar{M}_\pi^2$ and
$\frac{1}{2}(\hat{M}_{K^+}^2\!+\!\hat{M}_{K^0}^2)\!-\!\bar{M}_K^2$
are negligible compared to the uncertainties in the electromagnetic
self-energies, and the estimates given in (\ref{val_epsilon}) imply \cite{FLAG}
\beq
\bar{M}_\pi=134.8(3)\MeV\;,\qquad
\bar{M}_K=494.2(5)\MeV
\eeq
which means that our self-energy corrected, isospin-averaged pseudoscalar
input meson masses essentially agree with the PDG values of $M_{\pi^0}$ and
$\sqrt{\frac{1}{2}(M_{K^+}^2\!+\!M_{K^0}^2)}$, respectively.

\endcomment 

Since our lattice simulations are performed in the isospin symmetric limit
$m_d\!=\!m_u$ and do not account for electromagnetic interactions, the physical
input meson masses must be corrected for these effects.
The account of this as given by FLAG \cite{FLAG} is essentially a refined
version of the analysis presented by MILC \cite{Aubin:2004fs} some time ago.
The bottom line is that in the framework mentioned above one should use the
input masses $M_\pi=134.8(3)\MeV, M_K=494.2(5)\MeV$, which means
that the electromagnetically corrected, isospin-averaged pseudoscalar input
meson masses essentially agree with the PDG values of $M_{\pi^0}$ and
$\sqrt{\frac{1}{2}(M_{K^+}^2\!+\!M_{K^0}^2)}$, respectively.


\section{Simulation parameters}


\begin{table}[!p]
\vspace*{-8mm}
\centering
\begin{tabular}{|lcrcrcc|}
\hline
$\;\be$ & $am_{ud}^\mr{bare}$ & $am_s^\mr{bare}$ & volume & \#\,traj.\ & $a\Mpi$ & $\Mpi L$\\
\hline
\multirow{12}{*}{3.31}
 & -0.07000 & -0.0400 & $16^3\times 32$ & 1650 & 0.3530(12) & 5.61 \\
 & -0.09000 & -0.0400 & $24^3\times 48$ & 1600 & 0.2083(08) & 5.00 \\
 & -0.09300 & -0.0400 & $24^3\times 48$ & 4350 & 0.1775(06) & 4.30 \\
 & -0.09300 & -0.0400 & $32^3\times 48$ & 2500 & 0.1771(05) & 5.65 \\
 & -0.09530 & -0.0400 & $32^3\times 48$ & 1225 & 0.1500(13) & 4.81 \\
 & -0.09756 & -0.0400 & $32^3\times 48$ & 2600 & 0.1202(11) & 4.00 \\
 & -0.09900 & -0.0400 & $48^3\times 48$ & 1700 & 0.0887(06) & 4.26 \\
 & -0.09933 & -0.0400 & $48^3\times 48$ & 1240 & 0.0804(13) & 3.94 \\
 \cline{2-7}
 & -0.09000 & -0.0440 & $24^3\times 64$ & 1065 & 0.2024(10) & 4.86 \\
 & -0.09300 & -0.0440 & $32^3\times 64$ & 1030 & 0.1717(08) & 5.50 \\
 & -0.09530 & -0.0440 & $32^3\times 64$ & 1300 & 0.1457(09) & 4.66 \\
\hline
\multirow{12}{*}{3.5}
 & -0.04800 & -0.0023 & $32^3\times 64$ & 1500 & 0.1354(06) & 4.33 \\
 \cline{2-7}
 & -0.02500 & -0.0060 & $16^3\times 32$ &12000 & 0.2898(07) & 4.62 \\
 & -0.03100 & -0.0060 & $24^3\times 48$ & 3000 & 0.2535(05) & 6.07 \\
 & -0.03600 & -0.0060 & $24^3\times 48$ & 1800 & 0.2250(08) & 5.41 \\
 & -0.04100 & -0.0060 & $24^3\times 48$ & 4000 & 0.1921(05) & 4.61 \\
 & -0.04370 & -0.0060 & $24^3\times 48$ & 3900 & 0.1725(04) & 4.13 \\
 & -0.04900 & -0.0060 & $32^3\times 64$ & 1100 & 0.1212(08) & 3.90 \\
 & -0.05294 & -0.0060 & $64^3\times 64$ & 1100 & 0.0613(06) & 3.92 \\
 \cline{2-7}
 & -0.04100 & -0.0120 & $24^3\times 64$ & 1020 & 0.1884(08) & 4.52 \\
 & -0.04630 & -0.0120 & $32^3\times 64$ & 1065 & 0.1445(06) & 4.62 \\
 & -0.04900 & -0.0120 & $32^3\times 64$ & 1000 & 0.1174(06) & 3.76 \\
 & -0.05150 & -0.0120 & $48^3\times 64$ & 1200 & 0.0835(07) & 4.01 \\
\hline
\multirow{09}{*}{3.61}
 & -0.02000 &  0.0045 & $32^3\times 48$ & 2100 & 0.1993(3) & 6.36 \\
 & -0.02800 &  0.0045 & $32^3\times 48$ & 3910 & 0.1488(4) & 4.75 \\
 & -0.03000 &  0.0045 & $32^3\times 48$ & 2000 & 0.1322(4) & 4.24 \\
 & -0.03121 &  0.0045 & $32^3\times 48$ & 2200 & 0.1211(2) & 3.87 \\
 & -0.03300 &  0.0045 & $48^3\times 48$ & 2100 & 0.1026(4) & 4.93 \\
 & -0.03440 &  0.0045 & $48^3\times 48$ & 1100 & 0.0864(4) & 4.15 \\
 \cline{2-7}
 & -0.03650 & -0.0030 & $64^3\times 72$ & 1004 & 0.0468(5) & 3.00 \\
 \cline{2-7}
 & -0.02000 & -0.0042 & $32^3\times 48$ & 1750 & 0.1969(4) & 6.30 \\
 & -0.03000 & -0.0042 & $32^3\times 48$ & 1450 & 0.1297(5) & 4.17 \\
\hline
\multirow{09}{*}{3.7}
 & -0.00500 &  0.0500 & $32^3\times64$ & 1000 & 0.2227(04) & 7.13 \\
 & -0.01500 &  0.0500 & $32^3\times64$ & 1170 & 0.1711(03) & 5.48 \\
 \cline{2-7}
 & -0.02080 &  0.0010 & $32^3\times64$ & 1150 & 0.1251(04) & 4.00 \\
 \cline{2-7}
 & -0.01500 &  0.0000 & $32^3\times64$ & 1115 & 0.1644(04) & 5.26 \\
 & -0.02080 &  0.0000 & $32^3\times64$ & 1030 & 0.1245(06) & 3.98 \\
 & -0.02540 &  0.0000 & $48^3\times64$ & 1420 & 0.0821(03) & 3.94 \\
 & -0.02700 &  0.0000 & $64^3\times64$ & 1045 & 0.0603(03) & 3.86 \\
 \cline{2-7}
 & -0.02080 & -0.0050 & $32^3\times64$ & 1405 & 0.1249(04) & 4.00 \\
 & -0.02540 & -0.0050 & $48^3\times64$ & 1320 & 0.0806(03) & 3.87 \\
\hline
\end{tabular}
\newline
{... to be continued on next page ...}
\vspace*{-2mm}
\end{table}
\begin{table}[!t]
\centering
{... continued from previous page ...}
\\
\begin{tabular}{|lcrcrcc|}
\hline
\multirow{06}{*}{3.8}
 & -0.01400 &  0.0300 &   $32^3\times 64$   & 1325 & 0.1242(5) & 3.97 \\
 & -0.01900 &  0.0300 &   $48^3\times 64$   & 1045 & 0.0830(4) & 3.99 \\
 \cline{2-7}
 & -0.00900 &  0.0000 &   $32^3\times 64$   & 2280 & 0.1523(4) & 4.87 \\
 & -0.01400 &  0.0000 &   $32^3\times 64$   & 1055 & 0.1249(5) & 4.00 \\
 & -0.01900 &  0.0000 &   $48^3\times 64$   & 1080 & 0.0836(4) & 4.01 \\
 & -0.02100 &  0.0000 & $64^3\!\times\!144$ & 1200 & 0.0598(2) & 3.83 \\
\hline
\end{tabular}
\caption{\label{tab:params}\sl
Overview of our $N_f\!=\!2\!+\!1$ simulations.
The scales at $\be=3.31,3.5,3.61,3.7,3.8$
are $a^{-1}=1.697(6),2.131(13),2.561(26),3.026(27),3.662(35)\GeV$,
respectively. Accordingly, the minimum pion mass per coupling
is $\Mpi=136(2),131(2),120(2),182(2),219(2)\MeV$.}
\end{table}

An overview of our $N_f\!=\!2\!+\!1$ QCD simulations is presented in
Tab.\,\ref{tab:params}.
For each ensemble we indicate the bare parameters, the lattice geometry, and
the ensemble length in $\tau\!=\!1$ units (after thermalization).
In addition, the pion mass for the given parameters (determined with a specific
choice of the fitting interval) is given.
Note that the quoted error is only statistical~-- a detailed account of our
procedure to keep track of the systematic uncertainties is given in Sec.\,14.
With Wilson fermions negative bare masses are not disturbing; after
renormalization they will evaluate to positive quark masses (see Sec.\,11).
We work with spatial volumes as large as $L^3\!\simeq\!(6\fm)^3$ and temporal
extents up to $T\!\simeq\!8\fm$.
Besides reducing finite-volume corrections and excited-state contaminations,
large (four-dimensional) volumes tend to decrease statistical uncertainties to
the same extent as increasing the number of trajectories (in a fixed volume)
would do.
For instance, in a $L^4$ box 1300 trajectories at $\Mpi L\!=\!4$ are
approximately equivalent to 4100 trajectories at $\Mpi L\!=\!3$.
With an HMC-type algorithm, the effort (at fixed pion mass) grows like $L^5$.
Nevertheless, in view of the increased algorithmic stability (see below),
choosing large four-dimensional volumes is a beneficial strategy.

The integrated autocorrelation times of the smeared plaquette and of the number
of conjugate gradient iterations in the HMC accept/reject step are at most
$O(10)$ trajectories.
Depending on this autocorrelation time, the gauge field after every fifth or
every tenth trajectory is stored as a configuration to be used for calculating
hadronic observables.

We put sources for the correlation functions on up to eight timeslices.
For the precise form of the meson and baryon interpolating operators see
e.g.\ \cite{Montvay:1994cy}.
To reduce the relative weight of excited states in correlation functions
Gaussian sources and sinks are used, with a radius of about $0.32\fm$, which
was found to be a good choice \cite{Durr:2008zz}.


\section{Algorithm stability}


\begin{figure}[!b]
\centering
\includegraphics[width=15.5cm]{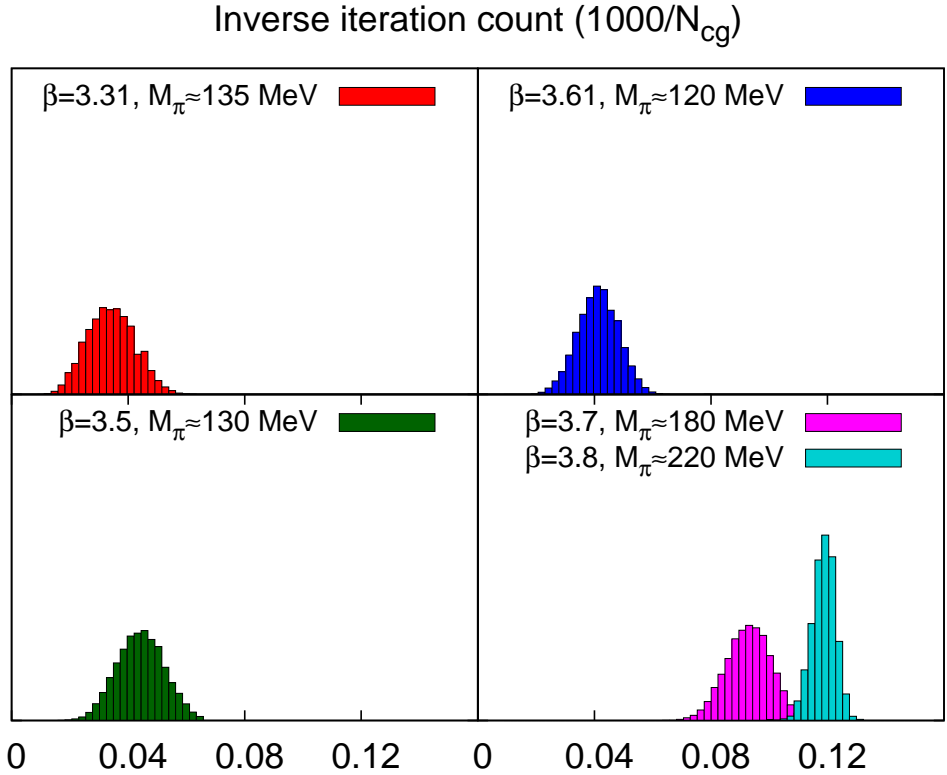}
\vspace*{-4mm}
\caption{\label{fig:stefan1}\sl
Histogram of the inverse iteration count ($N_\mr{CG}^{-1}$ of the lightest
pseudofermion) in the ensemble with the lightest quark mass per $\beta$. There
is no danger of a tail stretching out to zero.}
\end{figure}

\begin{figure}[p]
\vspace*{-6mm}
\centering
\includegraphics[width=13cm]{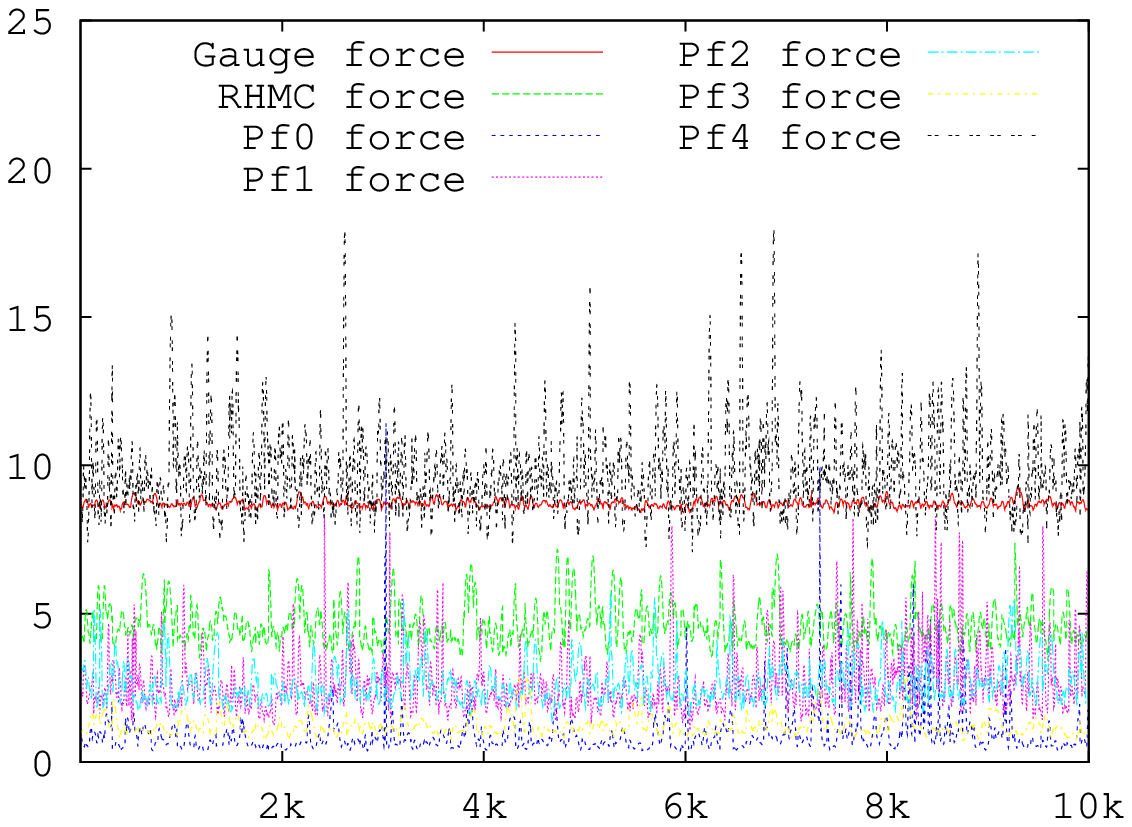}
\vspace*{-4mm}
\caption{\label{fig:stefan2}\sl
Evolution of the maximum of each MD force during the MD integration. 256 steps
correspond to one $\tau\!=\!1$ trajectory. Shown is one production stream of
our ``physical pion mass'' ensemble at $\beta\!=\!3.31$. The other streams with
the same parameters give a similar picture.}
\includegraphics[width=13cm]{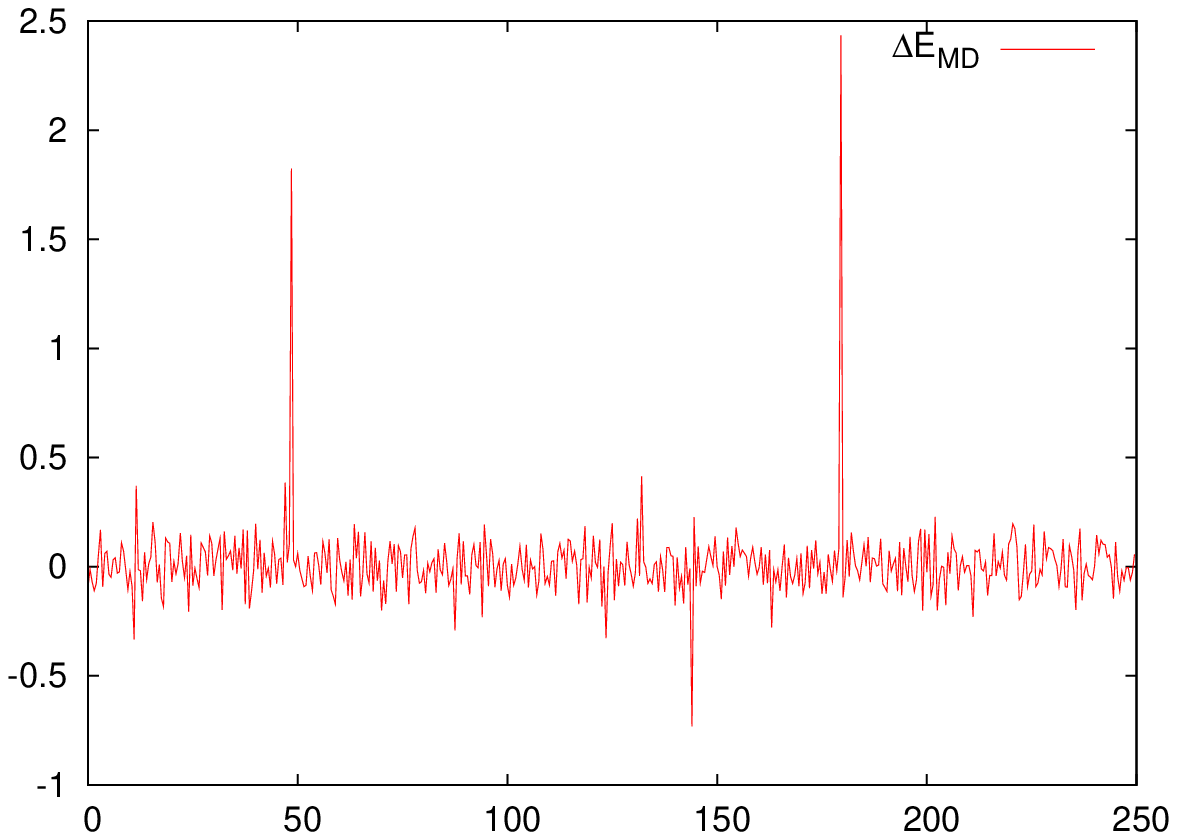}
\vspace*{-4mm}
\caption{\label{fig:stefan3}\sl
Evolution of the energy violation over 250 units of MD time in the same
simulation as in Fig.\,\ref{fig:stefan2}. All other
simulations show similar or smaller energy violations.}
\vspace*{-4mm}
\end{figure}

To detect potential instabilities of the HMC algorithm, different stability
tests need to be performed.
A rather complete battery of such tests was described in \cite{Durr:2008rw}.
The pion masses used in this work are considerably smaller than those
encountered in \cite{Durr:2008rw,Durr:2008zz}.
For this reason, we repeat the relevant stability tests for the smallest-mass
ensemble at each $\beta$, in particular for those with the physical pion mass.

With $D$ the Wilson or clover operator, the spectrum of $A\!=\!D\dag D$ has no
strict lower bound, i.e.\ the operator $A$ is positive semi-definite, but not
positive definite.
If one could integrate the HMC trajectories exactly, this would not cause any
problem, since an eigenvalue $\lambda$ of $A$ approaching zero would introduce
an unbounded back-driving force in the HMC, so that the exact zero would be
avoided.
In practice, the trajectories are generated with a finite step-size integrator.
Therefore, a very small $\lambda_\mr{min}$ in the MD evolution may experience a
smaller back-driving force than it would in an exact evolution scheme, and this
may trigger an instability.

If a particularly small eigenvalue appears during the molecular dynamics
(MD) evolution, the solver in the MD force calculation will require more
iterations to arrive at its target precision.
More precisely, the inverse of the iteration count $N_\mr{CG}$ is closely
related to the smallest eigenvalue of $A$.
In a given ensemble, $N_\mr{CG}^{-1}$ shows an approximately Gaussian
distribution \cite{Durr:2008rw}.
As long as its median is away from zero by several standard deviations, the
simulation is deemed safe \cite{DelDebbio:2005qa,Durr:2008rw}.
In Fig.\,\ref{fig:stefan1} we show the ``worst case scenario'', i.e.\ the
situation for the smallest quark masses in our set of ensembles.
As one can see, even for pion masses as small as $120\!-\!135\MeV$ the inverse
iteration count and hence the spectrum is bounded away from zero.

Alternatively, one can monitor the magnitude of the various contributions to
the MD force in the MD evolution.
This is done in Fig.\,\ref{fig:stefan2}, where the maximum (over space-time) of
each individual contribution to the total MD force during the MD integration is
shown over a period of $10,000$ MD steps.
As usual, the maximum of each contribution to the MD force fluctuates.
However, it is important that the magnitude of these fluctuations is not too
large.
The more frequent spikes with large magnitude occur at any given MD step-size,
the lower is the HMC acceptance rate, to the point where the algorithm becomes
unstable.
For small pion masses and coarse lattice spacings, the situation becomes even
worse.
This is why we show the ensemble with our smallest pion mass (around the
physical pion mass) at the coarsest lattice spacing in Fig.\,\ref{fig:stefan2}.
As one can see there are no dangerously large fluctuations present.

Finally, it is good practice to monitor the violation $\Delta\!E_\mr{MD}$ of
the MD energy conservation.
In Fig.\,\ref{fig:stefan3} we show again the ``worst case scenario'', that is
the ensemble with our smallest pion mass at the coarsest lattice spacing.
As one can see, the typical $\Delta\!E_\mr{MD}$ in this simulation is small.
For most of our simulations the acceptance rate is above 90\%.
Since the acceptance probability is given by
$p_\mr{acc}\!=\!\min(1,\exp(-\Delta\!E_\mr{MD}))$, it is reasonable to use the
data accumulated in the monitoring of the MD energy violation to check that
$\<\exp(-\Delta\!E_\mr{MD})\>\!=\!1$ within errors.

In summary, because ({\it a}) our algorithm is free of dangerous fluctuations
of the clover eigenvalue spectrum, ({\it b}) there are no dangerous
fluctuations in the MD forces and ({\it c}) we therefore see that large
violations of the MD energy conservation are absent in the simulation
(resulting in high acceptance rates), we conclude that our setup is safe.


\section{Autocorrelation and ergodicity checks}


\begin{figure}[!tb]
\centering
\includegraphics[width=14.5cm]{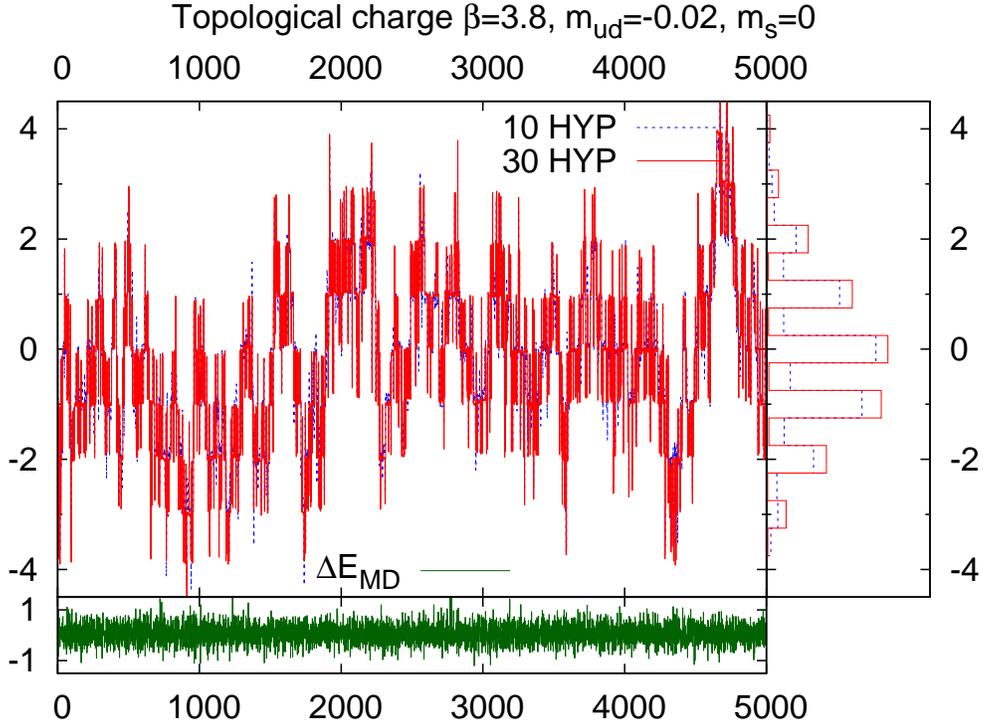}
\vspace*{-6mm}
\caption{\label{fig:hist-a}\sl
Topological charge history at our finest lattice spacing ($\be\!=\!3.8$
corresponds to $a^{-1}\!\sim\!3.7\GeV$) using two vigorous smearings (10 or 30
HYP steps) in the gluonic charge definition.}
\end{figure}

A known source of concern about HMC simulations in the regime of light quark
masses and/or small lattice spacings is whether the Markov chain manages to
sample configuration space sufficiently well, i.e.\ whether the algorithm is
(in practical terms) ergodic \cite{Alles:1996vn,Alles:1998jq,Schaefer:2009xx}.

We monitor two cheap gluonic quantities which are supposed to signal suspicious
behavior, if there is any.
The first one is the plaquette and/or Symanzik gauge action.
With the plaquette it makes sense to consider smeared varieties, too, i.e.\
$\mr{Re\,Tr}\,V_\mr{plaq}$ where $V$ is a smeared gauge link, as described in
Sec.\,2.
We find integrated autocorrelation times of such quantities to be at most
$O(10)$ in units of unit-length ($\tau\!=\!1$) trajectories.
The second quantity is the bare field-theoretical (global) topological charge
$q=a^4/(16\pi^2)\sum_{x,\mu<\nu}\mr{Tr}[F_{\mu\nu}(x)\tilde F_{\mu\nu}(x)]$
where $F_{\mu\nu}(x)$ is constructed from links which have undergone 10 or 30
steps of HYP smearing \cite{Hasenfratz:2001hp}.
The result for our finest lattice spacing which, according to
\cite{Alles:1996vn,Alles:1998jq,Schaefer:2009xx} represents the worst-case
scenario, is shown in Fig.\,\ref{fig:hist-a}.
First of all, the 10\,HYP and 30\,HYP charges are always close to each other.
Second, binning them with bin boundaries at $\mathbf{Z}/2\!+\!1/4$ yields a
clear abundance of integer centered bins over half-integer centered bins,
and this effect is more pronounced with the more vigorous smearing recipe.
Last but not least, the histogram of either charge is reasonably symmetric
after about $5000$ trajectories, and the integrated autocorrelation time is
$O(10)$ trajectories.
Since this is the highest $\beta$ simulated, we see no reason for concerns
about the (practical) ergodicity of our simulations.

One should keep in mind that the topological tunneling rate may depend
sensitively on the details of the action (e.g.\ whether Wilson, Symanzik or
Iwasaki glue is used, whether smeared or unsmeared links are used in the
fermionic part) and on the algorithm (e.g.\ the number of time scales
and the specific choice of Hasenbusch masses).


\section{$N_f\!=\!3$ scaling test for hadron masses}


\begin{figure}[!tb]
\includegraphics[width=7.9cm]{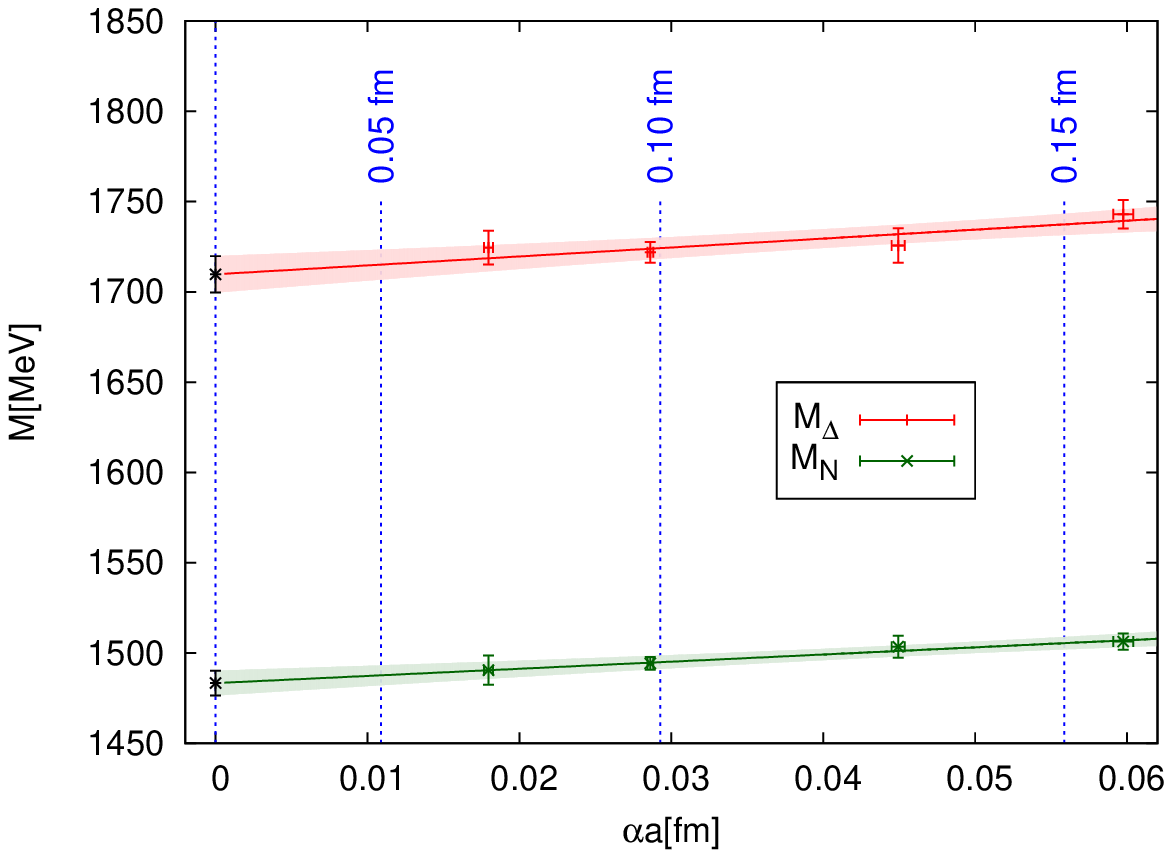}
\includegraphics[width=7.9cm]{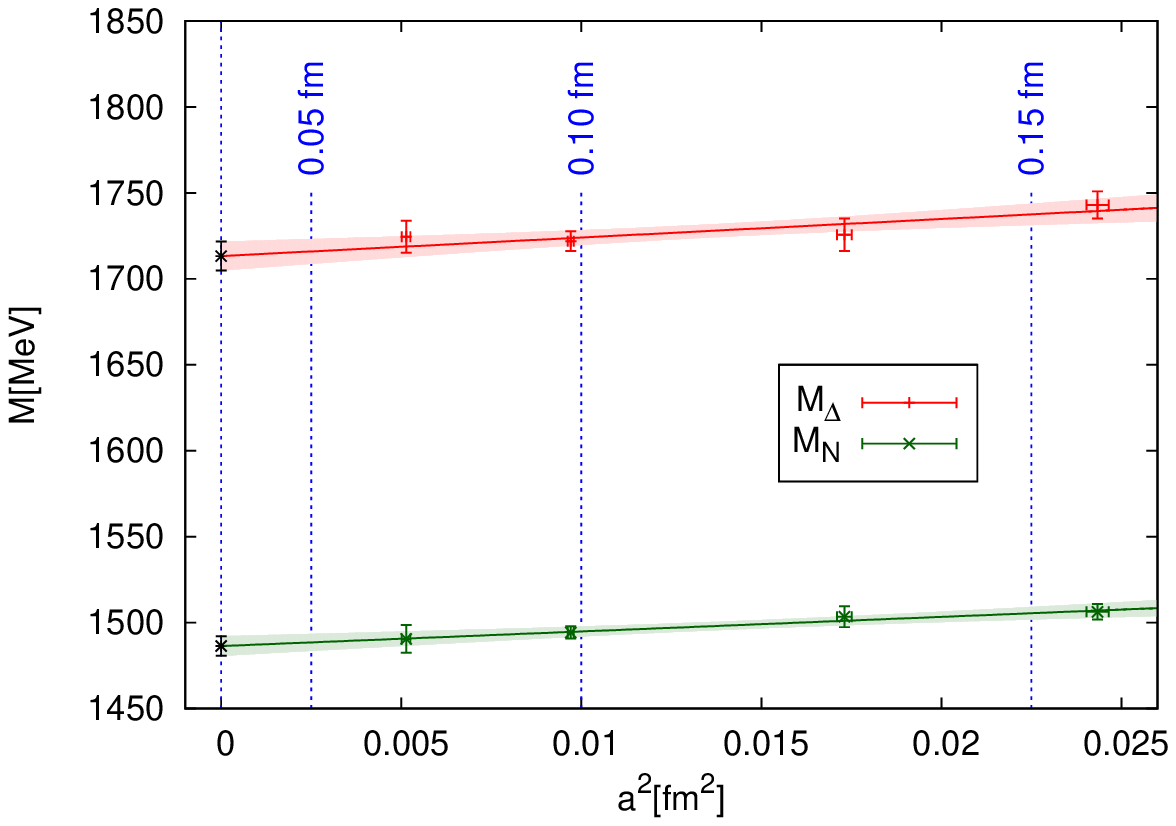}
\vspace*{-4mm}
\caption{\label{fig:scaling}\sl
Scaling of the nucleon and delta mass, at fixed $\Mpi/M_\rho=0.67$, versus
$\alpha a$ and $a^2$.}
\end{figure}

Since the link smearing of the 2HEX action used in the present work differs
from the smearing used in \cite{Durr:2008rw,Durr:2008zz}, we decided to repeat
the entire scaling test, as presented in \cite{Durr:2008rw}, in all its detail
for the new action.

To this end, we run a number of $N_f\!=\!3$ simulations at various lattice
spacings and various $\Mpi/M_\rho$ ratios.
For each $\be$ we interpolate the (common) octet and decuplet baryon mass,
i.e.\ $M_N/M_\rho$ and $M_\Delta/M_\rho$, to the point where $\Mpi/M_\rho$
assumes the value 0.67.
The latter value coincides with
$[2(\Mka^\mr{phys})^2-(\Mpi^\mr{phys})^2]^{1/2}/M_\phi^\mr{phys}$,
hence providing a way to tune to a quark mass which roughly corresponds to
the physical strange quark mass.
The results for $M_N/M_\rho$ and $M_\Delta/M_\rho$ at this interpolation point
are then extrapolated, linearly in $\alpha a$ and $a^2$, to the continuum.
Throughout this report, $\alpha\!=\!g^2/(4\pi)$ denotes the strong coupling
constant.
For $g^2$ we use the perturbative values, at our lattice spacings, at 4-loop
order (see below).
The result of this procedure is shown in Fig.\,\ref{fig:scaling}.
Three points are worth emphasizing.
First of all, the data are consistent with either scaling hypothesis over
a large range of lattice spacings (out to $a\!\simeq\!0.15\fm$), with a slight
preference for $O(a^2)$ over $O(\alpha a)$ scaling, and this suggests that our
tree-level value of $c_\mr{SW}$ (see Sec.\,2 for the definition and details) is
close to the nonperturbative value (which is not known for our action).
This finding is in accordance with the results of \cite{Capitani:2006ni}.
Next, the continuum extrapolated values shown in Fig.\,\ref{fig:scaling} are in
perfect agreement with the continuum extrapolated baryon masses found in
\cite{Durr:2008rw} with a different action.
Last but not least, the slope in either panel of Fig.\,\ref{fig:scaling} is
small%
\footnote{The deviation of the result on the coarsest lattice from the
continuum is 2.0\% at most [$\Delta$ with $O(\alpha a)$ ansatz].},
and an action which shows generically a flat slope in scaling quantities is
useful for obtaining precise predictions in the continuum.

In summary we find that both the 6stout action used in
\cite{Durr:2008rw,Durr:2008zz} and the 2HEX action used in the present work
exhibit small cut-off effects on standard hadron masses over a broad range of
lattice spacings.


\section{Finite volume corrections}


\begin{figure}[!tb]
\centering
\includegraphics[width=13cm]{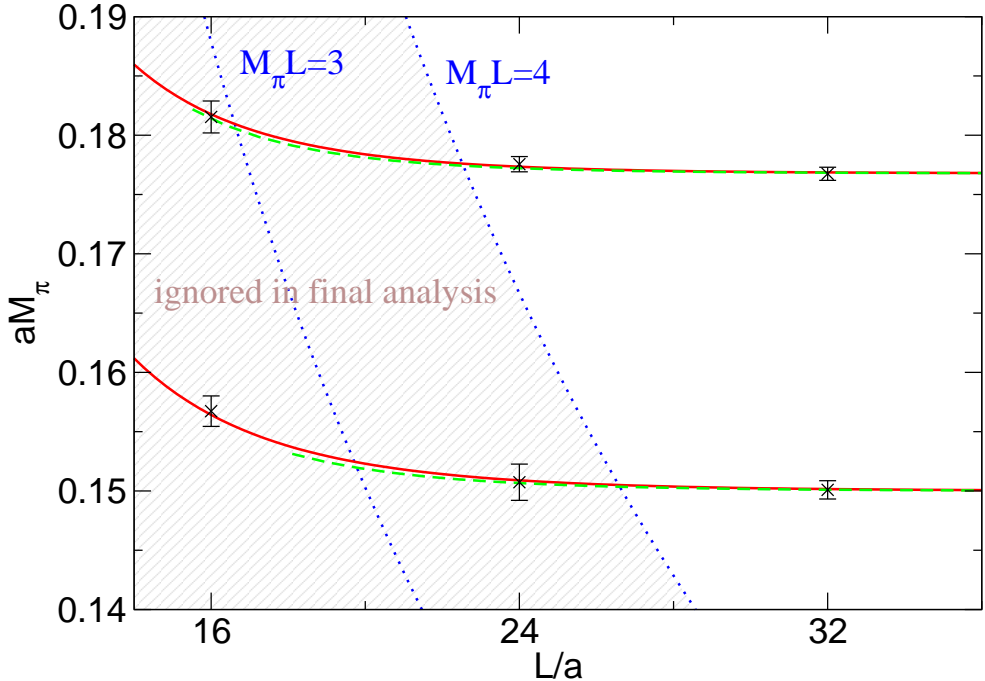}
\vspace{-3mm}
\caption{\label{fig:finvol}\sl
Dedicated finite-volume analysis at $\be\!=\!3.31$, with
$\Mpi\!\simeq\!250\MeV$ (lower set of data) and $\Mpi\!\simeq\!300\MeV$ (upper
set). Results are compared to the prediction from Chiral Perturbation Theory.
The fit to (\ref{GLFV}) is shown by solid red curves and the prediction of ChPT
\cite{Colangelo:2005gd} is the green set of dashed curves. The steep dotted
lines indicate the boundaries $\Mpi L\!=\!3$ and $\Mpi L\!=\!4$.}
\end{figure}

For a fixed set of bare parameters, $\be,m_{ud},m_s$, energies and matrix
elements of hadronic states depend on the spatial size $L$ of the lattice.
Typically, the finite volume tends to increase the effective mass and to
decrease the decay constant, relative to their infinite volume counterparts.
As a first step it is important to assess by how much such effects would affect
the data.
For the theoretical treatment of these finite-volume effects it makes a
difference whether the state considered is stable under strong interactions
(despite the fact that, in a finite volume, the energy spectrum is discrete
and all states are stable).
The respective frameworks have been established by L\"uscher, both for stable
\cite{Luscher:1985dn,Luscher:1986pf} and unstable
\cite{Luscher:1990ux,Luscher:1991cf} states.
They allow us to quantify and eventually correct for finite-volume effects in a
self-consistent manner (i.e.\ in a way in which only the results of our
calculations and axioms of quantum field theory are used).

The structure of these corrections is most transparent for the case of a pion
at 1-loop order in Chiral Perturbation Theory (ChPT)
\cite{Gasser:1986vb,Gasser:1987ah,Gasser:1987zq}.
Up to higher order terms, the relative shift is
\beq
R_\pi(L)\equiv\frac{\Mpi(L)}{\Mpi}-1=
\mr{const}\cdot\Mpi^2\cdot\til{g}_1(\Mpi L)
\label{GLFV}
\eeq
where the shape function $\til{g}_1(x)$ has a well behaved expansion
in terms of a Bessel function of the second kind, which itself has a
large-$x$ expansion of the form
\bea
\til{g}_1(x)&=&\frac{24K_1(x)}{x}+\frac{48K_1(\sqrt{2}x)}{\sqrt{2}x} + ...
\\
K_1(x)&=&\Big(\frac{\pi}{2x}\Big)^{1/2}\exp(-x)\Big[1+\frac{3}{8x}+...\Big]
\eea
implying  that finite volume corrections are exponentially suppressed at large
$L$ \cite{Luscher:1985dn}.
Higher loop orders for $R_\pi(L)$ have been worked out in
\cite{Colangelo:2005gd}.
For completeness we mention that analytic results for finite volume corrections
of the nucleon are given in \cite{AliKhan:2003cu,Colangelo:2010ba}.
The second reference predicts that for physical quark masses and $L\!=\!5\fm$
box size (which is what we use in our smallest box at the physical mass point,
the one at $\be\!=\!3.61$, cf.\,Tab.\ref{tab:params}) the nucleon experiences
just a 2 permil finite-size shift.
The point is that ChPT predicts the numerical value of the coefficient
``coeff'' in (\ref{GLFV}).
In the chiral literature, the low-energy constants that enter ``coeff'' are
pinned down from experiment (at leading order it is $\Fpi$).

To avoid using external input, we decide to stay content with just using the
functional form of (\ref{GLFV}).
This is permissible, since the shape function $\til{g}_1(x)$ is just the free
Green's function of a massive scalar particle, summed over all spatial mirror
copies (due to periodic boundary conditions in the spatial directions)
\cite{Gasser:1986vb}, see also the discussion in \cite{Colangelo:2005gd}.
We find that we can establish a global fit to all of our data in various
volumes if we adjust the free coefficient in (\ref{GLFV}).
A similar conclusion is reached for other stable hadrons, albeit with a
different numerical value of the constant.
For the case of the pion, we test the fitting ansatz and the analytic
prediction \cite{Colangelo:2005gd} by comparing them to dedicated finite-volume
scaling runs, as shown in Fig.\,\ref{fig:finvol}.
Both the fit (full line) and the prediction from ChPT \cite{Colangelo:2005gd}
(dashed curve) agree with the data.
The latter prediction has a limited range, since ChPT becomes questionable in
boxes with $\Mpi L\!<\!3$ \cite{Colangelo:2005gd}.
It is important to emphasize that the data with $\Mpi L<4$ in
Fig.\,\ref{fig:finvol} have been generated to \emph{test} our treatment of
finite-volume effects, they do not enter the final analysis.

These results confirm our rule of thumb that simulations with $\Mpi L\geq4$
and/or $L\gsim 5\fm$ yield infinite-volume masses within statistical accuracy.
An overview of the expected size of $R_{\Mpi}$ in our simulations is given in
Fig.\,\ref{fig:landscape}.
In all of these points the mass correction is less than about 5 permil, and for
points close to $\Mpi^\mr{phys}$ (which dominate our analysis) it is even
smaller.
Nevertheless, we include these (tiny) shifts into our global analysis (cf.\
Sec.\,14).


\section{Chiral behavior of pion mass and decay constant}


\begin{figure}[!tb]
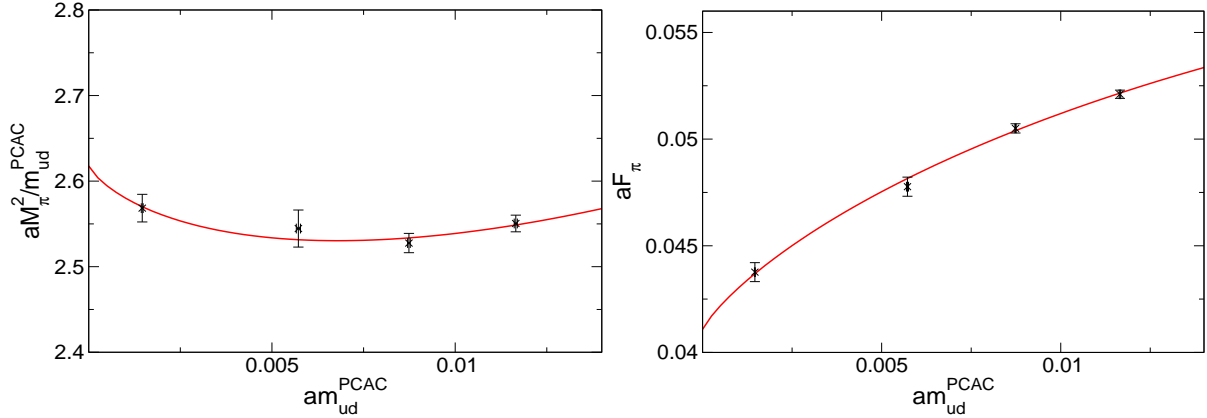

\includegraphics[width=7.9cm,height=5.5cm]{physpoint_prd.figs/combofit_4_b.eps}
\includegraphics[width=7.9cm,height=5.4cm]{physpoint_prd.figs/combofit_4_f.eps}
\caption{\label{fig:xpt}\sl
$\Mpi^2/m_{ud}^\mr{PCAC}$ (left) and $\Fpi$ (right) versus $m_{ud}^\mr{PCAC}$
(cf.\ Sec.\,11) for our 4 lightest ensembles at $\be\!=\!3.5$, at fixed
$am_s\!=\!-0.006$, which is close to $m_s^\mr{phys}$. A joint fit to the NLO
chiral ansatz (\ref{xpt_mpi}, \ref{xpt_fpi}) yields reasonable values of the
low-energy constants. Error bars are statistical.}
\end{figure}

To illustrate the quality of our results obtained in lattice QCD calculations
with physical or larger than physical values of the quark mass
$m_{ud}\!=\!(m_u\!+\!m_d)/2$, we briefly investigate here whether the $m_{ud}$
dependence of the pion mass and decay constant can be described by ChPT
\cite{Gasser:1983yg,Gasser:1984gg} in this range of quark masses.

To this end we compare our results for $\Mpi^2$ and $\Fpi$ versus $m_{ud}$ at
fixed (nearly physical) $m_s$ (cf.\ Tab.\,\ref{tab:params}) to the NLO
predictions of the $SU(2)$ framework.
The latter read \cite{Gasser:1983yg}
\bea
\Mpi^2&=&M^2\Big[1+{1\ovr2}x\log({M^2\ovr\Lambda_3^2})\Big]
\label{xpt_mpi}
\\
\Fpi  &=&\;F\;\,\Big[1-    x\log({M^2\ovr\Lambda_4^2})\Big]
\label{xpt_fpi}
\eea
with $x\!=\!M^2/(4\pi F)^2$ and $M^2\!=\!2Bm_{ud}$ a shorthand expression for
the light quark mass (up to the factor $2B$, with $B\!=\!\Sigma/F^2$).
The NNLO expressions can be found in \cite{Colangelo:2001df}.
In all of these expressions $F,\Sigma,B$ refer to the pion decay constant, the
absolute value of the quark condensate and the condensate parameter in the
2-flavor chiral limit $m_{ud}\to0$ with $m_s$ held fixed.
The terms in the square brackets proportional to $x^0,x^1$ represent the
tree-level and 1-loop contributions respectively, and $\Lambda_3,\Lambda_4$
encode low-energy constants (LECs).

In Fig.\,\ref{fig:xpt} the quantities $\Mpi^2/m_{ud}^\mr{PCAC}$ and $\Fpi$ are
plotted versus the PCAC quark mass $m_{ud}^\mr{PCAC}$ (see below) at an
intermediate lattice spacing ($\beta\!=\!3.5$), where we reach down to
$\Mpi\!\simeq\!130\MeV$ (cf.\ Tab.\,\ref{tab:params}).
We find that our results with $\Mpi\!<\!400\MeV$ can be jointly fitted with the
NLO chiral formulae (\ref{xpt_mpi}, \ref{xpt_fpi}), with acceptable
$\chi^2/\mr{d.o.f.}$ and reasonable values of the low-energy constants.
However, the extraction of Gasser-Leutwyler coefficients is beyond
the scope of this paper and will be left for future publications.


\section{RI/MOM renormalization of quark masses}


Our primary goal is to determine the average up-down quark mass
$m_{ud}=(m_u+m_d)/2$ and the strange quark mass $m_s$ at the physical mass
point in a ``continuum'' renormalization scheme, such as RI/MOM or RGI, using
first-principles lattice computations.

In a lattice study quark masses and the running coupling have a different
status than other observables, such as hadron masses and decay constants, since
they are input parameters to the simulation.
Consequently, one has to tune these parameters until the low-energy spectrum
of the theory agrees with experiment (cf.\ \cite{Durr:2008zz} and Sec.\,4),
before one may read them off from the results of the simulation.
To turn them into observables, one has to convert them from the original
cut-off scheme (which is specific to the gluon and quark action combination
used) to a scheme where the scale $\mu$ is not tied to the lattice spacing $a$.

The remainder of this section is organized as follows.
In 11.1 our ``ratio difference method'' for extracting quark masses in the
theory with Wilson fermions is explained, using standard terminology for the
renormalization and improvement coefficients.
It is important to notice that in the dynamical theory there is a subtlety in
the renormalization pattern, due to quark line disconnected diagrams
\cite{Gockeler:2004rp,Rakow:2004vj,Bhattacharya:2005rb}, but our ``ratio
difference method'' steers around this complication, as explained in 11.2.
In subsection 11.3 details of how we determine the flavor non-singlet scalar
renormalization constant $Z_S(\mu)$ via the Rome-Southampton RI/MOM method
\cite{Martinelli:1994ty} are given.
In 11.4 it is specified how we control the systematics that arise from the
dedicated $N_f\!=\!3$ computations needed in the RI/MOM procedure.
In 11.5 a summary is given.


\subsection{Ratio-difference method in a nutshell}

With Wilson-type fermions there are two options for obtaining the renormalized
quark mass.
On the one hand, one may start from the mass parameter $am^\mr{bare}$, as
present in the Lagrangian, and apply both additive and multiplicative
renormalization to build the VWI quark mass
\beq
m^\mr{VWI}=\frac{1}{Z_S}\,\Big[1-\frac{1}{2}b_Sam^\mr{W}+O(a^2)\Big]\,m^\mr{W}
\qquad\mbox{where}\qquad
m^\mr{W}=m^\mr{bare}-m^\mr{crit}
\label{def_vwi}
\eeq
and VWI means%
\footnote{Strictly speaking the vector Ward identity constrains only quark
mass differences. Below we will use $m^\mr{VWI}$ only in such differences, and
by doing so the dependence on $m^\mr{crit}$ will persist only in an
$O(a)$-suppressed term.}
``vector Ward identity''.
Here $Z_S\!=\!Z_S(\mu)$ denotes the lattice-to-continuum renormalization factor
of the scalar density (it depends on the chosen scheme and scale, e.g.\
$\MSbar$ and $\mu\!=\!2\GeV$), $b_S$ is an improvement coefficient (see below),
and $m^\mr{crit}$ specifies the additive mass renormalization, i.e.\ the bare
quark mass at which the pion becomes massless.

The other way to obtain the renormalized quark mass is as follows.
For a pseudoscalar meson made out of valence quarks $\psi_1,\psi_2$ with
Lagrangian masses $m_i^\mr{bare}$ or Wilson masses $m_i^\mr{W}$ ($i\!=\!1,2$),
respectively, the sum of the (unrenormalized) PCAC quark masses is defined as
\beq
m_1^\mr{PCAC}+m_2^\mr{PCAC}=\frac
{\sum_{\vec{x}}\<\bar\partial_\mu[A_\mu(x)+ac_A\bar\partial_\mu P(x)]O(0)\>}
{\sum_{\vec{x}}\<P(x)O(0)\>}
\label{mquark_PCAC}
\eeq
where $A_\mu$ and $P$ denote the axial current and the pseudoscalar density,
respectively, $O$ represents an arbitrary operator which couples to the meson
(usually $O\!=\!P$ to maximize the signal), and
$\bar\partial_\mu\!=\!(\partial_\mu^{}\!+\!\partial_\mu^*)/2$ is the symmetric
derivative with
$(\partial_\mu\phi)(x)\!=\!(\phi(x\!+\!a\hat\mu)\!-\!\phi(x))/(2a)$.
Then only a multiplicative renormalization is needed to form the (renormalized)
``AWI quark mass''
\beq
m^\mr{AWI}={Z_A\ovr Z_P}\,
\frac{1+b_A am^\mr{W}+O(a^2)}{1+b_P am^\mr{W}+O(a^2)}\,m^\mr{PCAC}
\label{def_awi}
\eeq
where AWI stands for ``axial Ward identity''.
Here $Z_A$ and $Z_P\!=\!Z_P(\mu)$ denote the lattice-to-continuum
renormalization factor of the axial current and the pseudoscalar density,
respectively.

The coefficients $b_S,b_A,b_P,c_A$ in (\ref{def_vwi}\,-\,\ref{def_awi}) are
part of the improvement program.
If properly set, $O(a^2)$ scaling of phenomenological quantities can be
achieved, but they may be set to zero if one is content with $O(a)$ scaling.
We use (\ref{def_vwi}\,-\,\ref{def_awi}) with tree-level values
of the improvement coefficients, that is $b_S\!=\!b_A\!=\!b_P\!=\!1$ and
$c_A\!=\!0$ .
Formally, our results thus show cut-off effects proportional to $\alpha a$, but
in the scaling tests presented above cut-off effects proportional to $a^2$ seem
to be numerically dominant.
At this point we cannot anticipate whether a similar statement holds true for
renormalized quark masses, and we shall thus consider both possibilities (i.e.\
leading cut-off effects proportional to $\alpha a$ or $a^2$).
In any case the difference (in a given scheme, at a given $\mu$) scales away
with $a\!\to\!0$, hence $m^\mr{AWI}\!=\!m^\mr{VWI}$ in the continuum.

In principle, $m_s^\mr{phys}$ and $m_{ud}^\mr{phys}$ may be determined using
either definition of the quark mass, but in practice it proves beneficial to
combine the specific advantages of the VWI and AWI approaches.
Let us assume, for a moment, that we were to set all improvement coefficients
to zero.
Since the Lagrangian quark mass $m^\mr{bare}$ is an exact input quantity which,
after a universal shift has been applied, multiplicatively renormalizes with
the unproblematic scalar density [cf.\ (\ref{def_vwi})], it is natural to use
$m^\mr{W}$ for quark mass \emph{differences}, where the additive
renormalization term $m^\mr{crit}$ drops out.
On the other hand, the PCAC quark mass $m^\mr{PCAC}$ is perfectly suited to
measure quark mass \emph{ratios}, since in the ratio the multiplicative
renormalization constants cancel [cf.\ (\ref{def_awi})].
It is thus natural to measure the difference $m_s\!-\!m_{ud}$ via the
Wilson or Lagrangian mass difference $d\!\equiv\!am^\mr{W}_s\!-\!am^\mr{W}_{ud}
\!=\!am^\mr{bare}_s\!-\!am^\mr{bare}_{ud}$ and
the ratio $m_s/m_{ud}$ via the PCAC mass ratio
$r\equiv m^\mr{PCAC}_s/m^\mr{PCAC}_{ud}$.
In this case one obtains the renormalized masses through
\beq
am_{ud}^\mr{scheme}=\frac{1}{Z_S^\mr{scheme}}\frac{d}{r-1}
\;,\qquad
am_s^\mr{scheme}=\frac{1}{Z_S^\mr{scheme}}\frac{rd}{r-1}
\label{ratio_diff_simple}
\eeq
and we shall refer to this strategy as the ``ratio-difference method''.
The renormalization scheme will be specified below.
In practice, things are slightly more involved, as we intend to maintain
tree-level improvement.
Setting $c_A\!=\!0$ and $b_A\!=\!b_P\!=\!1$ does not change anything in
(\ref{mquark_PCAC}, \ref{def_awi}), but having a quadratic term in
(\ref{def_vwi}) through $b_S\!=\!1$ means that the difference
$m_1^\mr{W}\!-\!m_2^\mr{W}$ does no longer coincide with
$m_1^\mr{bare}\!-\!m_2^\mr{bare}$.
In the next subsection we will show that even with improvement, the
renormalized quark masses are given by (\ref{ratio_diff_simple}) with
$d\to d^\mr{imp}, r\to r^\mr{imp}$, where the latter quantities are defined in
(\ref{def_dimp}, \ref{def_rimp}).


\subsection{Ratio-difference method in full QCD with improvement}

In the dynamical theory, the renormalization pattern of quark masses is
slightly more involved than the familiar equations (\ref{def_vwi},
\ref{def_awi}) suggest \cite{Bhattacharya:2005rb}, but it turns out that our
``ratio difference method'' gets rid of these complications and the final
relation is unchanged.

We now discuss how the findings of \cite{Bhattacharya:2005rb} apply to our
method, using their notation, except that we do not use a ``hat'' to denote
renormalized quantities, since they will come with a superscript ``VWI'' or
``AWI'', just as in the previous subsection.
Equations (26,\,48) of \cite{Bhattacharya:2005rb} read
\beq
m_j^\mr{VWI}=\frac{1}{Z_S}m_j^\mr{W}
\Big[
1-\frac{1}{2}b_Sam_j^\mr{W}-\bar{b}_Sa\tr(M)+O(a^2)
\Big]
+...
\label{def_vwi_bhatta}
\eeq
\beq
m_j^\mr{AWI}=\frac{Z_A}{Z_P}m_j^\mr{PCAC}
\Big[
1+(b_A\!-\!b_P)am_j^\mr{W}+(\bar{b}_A\!-\!\bar{b}_P)a\tr(M)+O(a^2)
\Big]
\label{def_awi_bhatta}
\eeq
where $Z_J$ is the flavor \emph{non-singlet} renormalization constant
($J\!=\!S,A,P$) , and $b_J\!=\!1\!+\!O(\alpha)$, $\bar{b}_J\!=\!O(\alpha^2)$,
$c_A\!=\!O(\alpha)$ denote improvement coefficients (which now depend on $N_f$).
Finally, $m_j^\mr{W}\!=\!m_j^\mr{bare}\!-\!m^\mr{crit}$, with $m^\mr{crit}$
defined as the $N_f\!=\!3$ critical mass (i.e.\ in the \emph{unitary}
direction), $m^\mr{PCAC}$ just as in (\ref{mquark_PCAC}), and the ellipses
in (\ref{def_vwi_bhatta}) denote terms which depend on the quark masses only
through $\tr(M),\tr(M^2),\tr^2(M)$, where $M$ is the (flavor diagonal) quark
mass matrix.
The new feature of formulas (\ref{def_vwi_bhatta}, \ref{def_awi_bhatta}) is the
terms proportional to $m_j$ times $\tr(M)\!=\!\sum_fm_f^\mr{W}$.
These terms make the renormalized quark mass of flavor $j$ depend on all other
quark masses, too.
Evidently, these terms come from quark loops in the functional determinant, and
the perturbative expansion of the new improvement coefficients
$\bar{b}_S,\bar{b}_A,\bar{b}_P$ shows that they start out at order $g_0^4$,
which means that they come through two-loop effects (one quark loop and a gluon
loop which attaches it to the quark line whose renormalization is studied).

Upon considering the difference of two VWI masses and the ratio of two AWI
masses
\beq
m_j^\mr{VWI}-m_k^\mr{VWI}=\frac{1}{Z_S}(m_j^\mr{W}-m_k^\mr{W})
\Big[
1-\frac{1}{2}b_Sa(m_j^\mr{W}\!+\!m_k^\mr{W})-\bar{b}_Sa\tr(M)+O(a^2)
\Big]
\eeq
\beq
\frac{m_j^\mr{AWI}}{m_k^\mr{AWI}}=\frac{m_j^\mr{PCAC}}{m_k^\mr{PCAC}}
\Big[
1+(b_A\!-\!b_P)a(m_j^\mr{W}\!-\!m_k^\mr{W})+O(a^2)
\Big]
\eeq
the term proportional to $a\tr(M)$ disappears from the second relation, and the
term proportional to $b_S$ involves only the sum of the Wilson masses.
Applying these formulas to $m_s$ and $m_{ud}$ in $N_f\!=\!2\!+\!1$ QCD, with
$d\!\equiv\!am_s^\mr{bare}\!-\!am_{ud}^\mr{bare}$ and
$r\!\equiv\!m_s^\mr{PCAC}/m_{ud}^\mr{PCAC}$ defined as before, one has
\beq
am_s^\mr{VWI}-am_{ud}^\mr{VWI}=\frac{1}{Z_S}\,d\,
\Big[
1-\frac{1}{2}b_Sa(m_s^\mr{W}\!+\!m_{ud}^\mr{W})-
\bar{b}_Sa(m_s^\mr{W}\!+\!2m_{ud}^\mr{W})+O(a^2)
\Big]
\eeq
\beq
\frac{m_s^\mr{AWI}}{m_{ud}^\mr{AWI}}=r\,
\Big[
1+(b_A\!-\!b_P)a(m_s^\mr{W}\!-\!m_{ud}^\mr{W})+O(a^2)
\Big]
\eeq
where we have used $d$ and $r$ only in the leading term, so far.
The point is that
\beq
am_s^\mr{W}+am_{ud}^\mr{W}=
(am_s^\mr{W}\!-\!am_{ud}^\mr{W})\,
\frac{m_s^\mr{W}/m_{ud}^\mr{W}+1}{m_s^\mr{W}/m_{ud}^\mr{W}-1}\simeq
d\,\frac{r+1}{r-1}
\eeq
\beq
am_s^\mr{W}\!+\!2am_{ud}^\mr{W}=
(am_s^\mr{W}\!-\!am_{ud}^\mr{W})\,
\frac{m_s^\mr{W}/m_{ud}^\mr{W}+2}{m_s^\mr{W}/m_{ud}^\mr{W}-1}\simeq
d\,\frac{r+2}{r-1}
\eeq
where the approximately equal sign means ``up to terms of order $O(a^2)$''.
Accordingly, we can express the difference of the VWI
masses and the ratio of the AWI masses through $d$ and $r$ as
\beq
am_s^\mr{VWI}-am_{ud}^\mr{VWI}=\frac{1}{Z_S}\,d^\mr{imp}
\quad,\qquad
\frac{m_s^\mr{AWI}}{m_{ud}^\mr{AWI}}=r^\mr{imp}
\eeq
where $d^\mr{imp}$ and $r^\mr{imp}$ are defined as
\beq
d^\mr{imp}=d\,
\Big[
1-\frac{1}{2}b_Sd\frac{r+1}{r-1}-\bar{b}_Sd\frac{r+2}{r-1}+O(a^2)
\Big]
\label{def_dimp}
\eeq
\beq
r^\mr{imp}=r\,
\Big[
1+(b_A\!-\!b_P)d+O(a^2)
\Big]
\;.
\label{def_rimp}
\eeq
In total this means that one finds $am_{ud}^\mr{scheme}$ and $am_s^\mr{scheme}$
via (\ref{ratio_diff_simple}) with $d\to d^\mr{imp}$, $r\to r^\mr{imp}$.

In our analysis, the tree-level improvement strategy makes all subleading terms
in the square brackets of (\ref{def_dimp}, \ref{def_rimp}) disappear, except
for the one proportional to $b_S$ (with $b_S\!=\!1$).


\subsection{Determination of the scalar RI/MOM renormalization factor}

Having laid out our overall strategy for obtaining the renormalized quark
masses $m_{ud}^\mr{phys}(\mu)$, $m_s^\mr{phys}(\mu)$ at the physical mass point
in a standard scheme at a given scale $\mu$, we now give details of how we
compute the single renormalization factor needed, $Z_S(\mu)$.
We implement the nonperturbative Rome-Southampton method which defines the
regularization-independent (RI/MOM) scheme \cite{Martinelli:1994ty}, with
several practical refinements (see below).
In the terminology of \cite{Gockeler:2004rp,Rakow:2004vj,Bhattacharya:2005rb}
the result is the non-singlet renormalization factor $Z_S^\mr{NS}(\mu)$.
In the RI/MOM scheme the running of $Z_S(\mu)$ is known perturbatively to
4-loop order \cite{Chetyrkin:1999pq}.
However, this is only relevant for the conversion to other schemes, e.g.\
$\MSbar$ at $\mu\!=\!2\GeV$.
Our main result, $m_{ud}$ and $m_s$ in the RI/MOM scheme at $\mu\!=\!4\GeV$
is derived without reference to perturbation theory.

In the RI/MOM scheme, renormalization conditions are defined in Landau gauge
and require the forward, truncated quark Green's function of an operator to be
equal to its tree-level value at a Euclidean four-momentum $p$, whose magnitude
is chosen to be the renormalization scale.
Given a quark bilinear operator $O_{12}^\Gamma=\bar\psi_2\Gamma\psi_1$, where
$\psi_1$ and $\psi_2$ are mass-degenerate quark fields and $\Gamma$ is a Dirac
matrix, the relevant Green's function is
\beq
\Lambda_\Gamma(p)\equiv
\<S(p)\>^{-1}\left\{\int\,d^4xd^4y\,e^{\mr{i}p(x\!-\!y)}\,
\langle\psi_2(y)O_{12}^\Gamma(0)\bar\psi_1(x)\rangle\right\}\<S(p)\>^{-1}
\;.
\eeq
In this equation, $S(p)$ is the propagator of quark flavors 1 and 2.
Now, defining a projector $P_\Gamma$ such that
$\mathrm{tr}\{P_\Gamma\Gamma\}=1$ (the trace is over spin$\times$color), the
renormalization condition reads
\beq
Z_\Gamma(\mu)=Z_\psi(\mu)\,/\,\Gamma_\Gamma(p)\vert_{p^2=\mu^2}
\label{RI_condition}
\;
\eeq
where $Z_\psi$ is the wavefunction renormalization factor and
\beq
\Gamma_\Gamma(p)\equiv
\mathrm{tr}\{P_\Gamma\Lambda_\Gamma(p)\}
\;.
\eeq

\begin{figure}[tb]
\centering
\includegraphics[width=13cm]{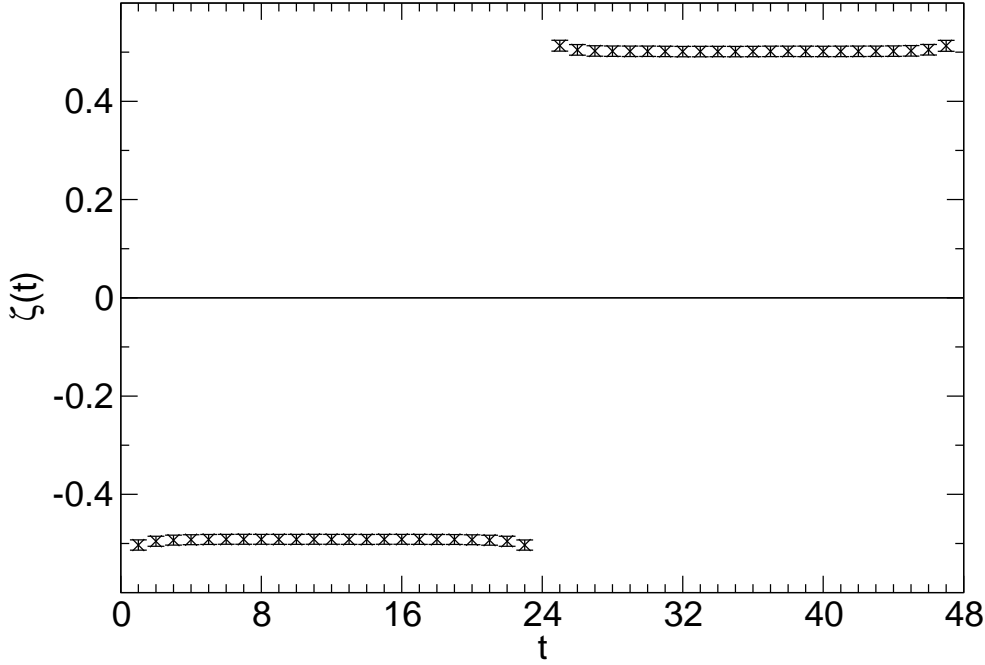}
\caption{\label{fig:zeta}\sl
The ratio $\zeta(t)$ as defined in (\ref{def_zeta}). The gauge coupling in this
$N_f\!=\!3$ run is $\be\!=\!3.61$, the quark mass is $am\!=\!-0.0045$. This
procedure yields a stable plateau for $Z_V$.}
\end{figure}

\begin{figure}[tb]
\centering
\includegraphics[width=13cm]{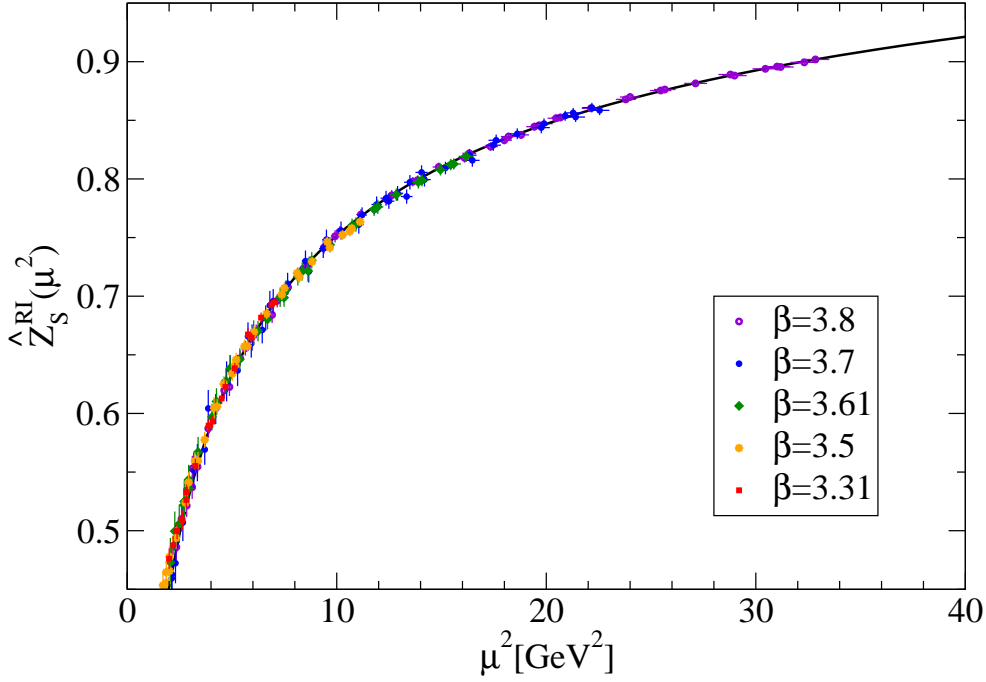}
\vspace{-2mm}
\caption{\label{fig:ch_figr}\sl
Renormalization factors $Z_{S,\be}^\mr{RI}(\mu^2)$ as a function of the bosonic
momentum squared. For each $\be$ momenta $\mu\!\leq\!\pi/(2a)$ are included.
The data from the coarser lattices have been multiplied by a $\mu$-independent
factor to match those at $\be\!=\!3.8$. The solid line represents a Pade ansatz
where the 1-loop anomalous dimension is built in as a constraint.}
\end{figure}

To determine $Z_S$ from the RI/MOM condition (\ref{RI_condition}) with
$\Gamma\!=\!I$, one needs to know $Z_\psi$.
In the original publication \cite{Martinelli:1994ty} the procedure was
supplemented with a recipe to obtain $Z_\psi$ from the momentum dependence of
the trace of the inverse propagator.
Here we avoid the determination of this wave-function renormalization constant
all together, by calculating the ratio $Z_S(\mu)/Z_V$ via the RI/MOM procedure
and by combining it with $Z_V$ from the 3-point function with a vector-current
insertion.
In other words, on each ensemble we compute $Z_S(\mu)/Z_V$ using
\beq
\frac{Z_{S,\be,m}(\mu)}{Z_{V,\be,m}}=
\frac{\Gamma_V(p)}{\Gamma_S(p)}\bigg|_{\hat{p}^2=\mu^2}
\label{RI_VoverS}
\eeq
where the dependence on the coupling and the $N_f\!=\!3$ quark mass is
indicated with subscripts.
The bosonic momentum definition $\hat{p}_\nu\!=\!(2/a)\sin(ap_\nu/2)$ is used,
and a standard cylinder cut around hyperdiagonal momenta is applied
\cite{Leinweber:1998im}.
In addition, we determine $Z_V$ from the ratio
\beq
\zeta(t)\equiv
\frac{\langle P(T/2)V_4(t)\bar P(0)\rangle}{\langle P(T/2)\bar P(0)\rangle}
\label{def_zeta}
\eeq
where
\beq
P(t)=\sum_{\vec{x}}(\bar\psi_2\gamma_5\psi_1)(\vec{x},t)\;,\quad
\bar{P}(t)=\sum_{\vec{x}}(\bar\psi_1\gamma_5\psi_2)(\vec{x},t)\;,\quad
V_4(t)=\sum_{\vec{x}}(\bar\psi_1\gamma_4\psi_1)(\vec{x},t)
\eeq
and $T$ denotes the temporal extent of the lattice.
With tree-level improvement one has \cite{Bakeyev:2003ff,Bhattacharya:2005rb}
\beq
Z_{V,\be,m}\,(1+am^\mr{W})=[\zeta(t_1)-\zeta(t_2)]^{-1}
\quad\mbox{for}\quad
0\!<\!t_1\!<T/2\!<\!t_2\!<\!T
\label{improved_ZV}
\eeq
where $b_V\!=\!1, \bar{b}_V\!=\!0$ have been used, and Fig.\,\ref{fig:zeta}
shows the plateau from which we extract $Z_{V,\be,m}$.
Combining this factor with the result of (\ref{RI_VoverS}) yields
$Z_{S,\be,m}(\mu)$, much in the spirit of
\cite{Martinelli:1990ny,Martinelli:1993dq}.


\subsection{Controlling the systematics in the RI/MOM procedure}

RI/MOM is a mass independent scheme.
Applied to the numerical data for $Z_S^\mr{RI}(\mu)$ this means that we have to
extrapolate all three flavors to vanishing sea and valence quark mass.
For this reason, we have generated a series of dedicated $N_f\!=\!3$ lattices
(i.e.\ with three degenerate quarks), where the action
$S\!=\!S_g^\mr{Sym}\!+\!S_f^\mr{SW}$ and the couplings $\be\!=\!6/g^2$ are the
same as in the phenomenological ensembles.
The bare parameters and statistics of these runs are summarized in
Tab.\,\ref{tab:runs_nf3}.
The specifics of the extrapolation will be discussed below.

\begin{table}[tb]
\centering
\begin{tabular}{|cc|cc|cc|cc|cc|}
\hline
3.31 & $\!\!16^3\!\times\!32\!$ &
3.5  & $\!\!24^3\!\times\!48\!$ &
3.61 & $\!\!24^3\!\times\!48\!$ &
3.7  & $\!\!32^3\!\times\!64\!$ &
3.8  & $\!\!32^3\!\times\!64\!$ \\
\hline
-0.04 & 4780 & -0.006 & 2560 & -0.0045 & 4620 & -0.0060 & 1010&\,\,0.000&\,\,505 \\
-0.06 & 3320 & -0.010 & 3140 & -0.0085 & 3680 & -0.0085 & 1050 &  -0.004&\,\,635 \\
-0.07 & 2420 & -0.012 & 2580 & -0.0100 & 4140 & -0.0110 & 1020 &  -0.008&\,\,500 \\
-0.08 & 2500 & -0.020 & 2700 & -0.0200 & 3140 & -0.0140 & 1290 &  -0.012 &  1030 \\
      &      & -0.035 & 1090 & -0.0250 & 1230 & -0.0160 & 1020 &  -0.014 &  1000 \\
\hline
\end{tabular}
\caption{\label{tab:runs_nf3}\sl
Bare masses and number of trajectories of our dedicated $N_f\!=\!3$ simulations
for RI/ MOM renormalization. The $\be$-values are the same as in our
phenomenological runs, cf.\ Tab.\,\ref{tab:params}.}
\end{table}

\begin{figure}[tb]
\centering
\includegraphics[width=13cm]{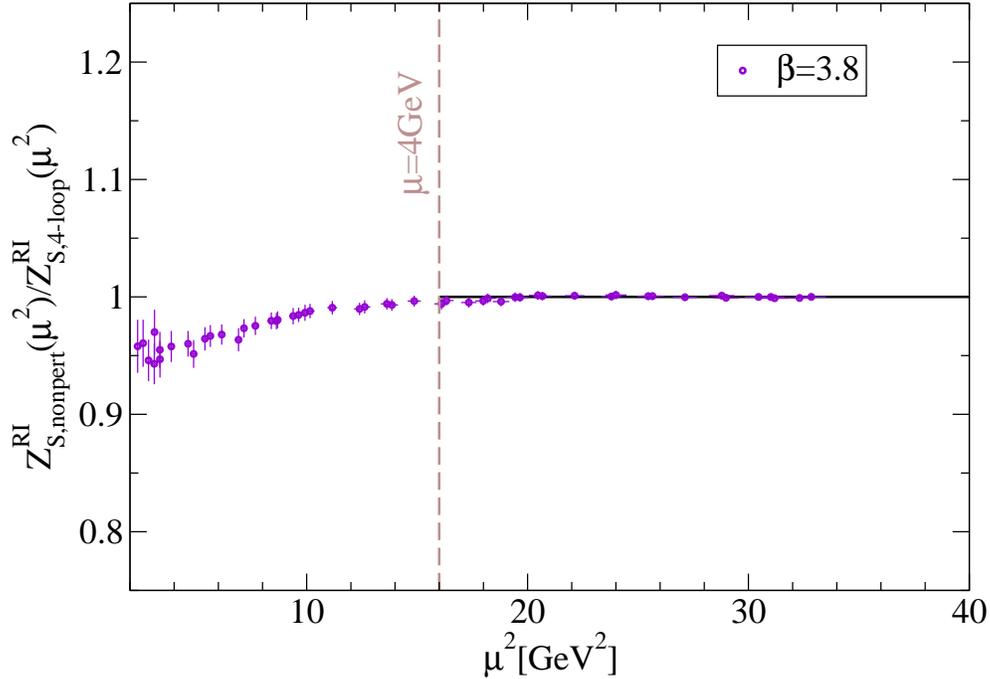}
\vspace{-2mm}
\caption{\label{fig:ch_figplat}\sl
Ratio of the nonperturbative $Z_S^\mr{RI}$ to the perturbative prediction at
4-loop level. The momentum range shown extends to $\mu_\mr{max}\!=\!\pi/(2a)$
at $\be\!=\!3.8$. For $\mu\!\geq\!4\GeV$ the data agree with the plateau within
errors.}
\end{figure}

\begin{figure}[p]
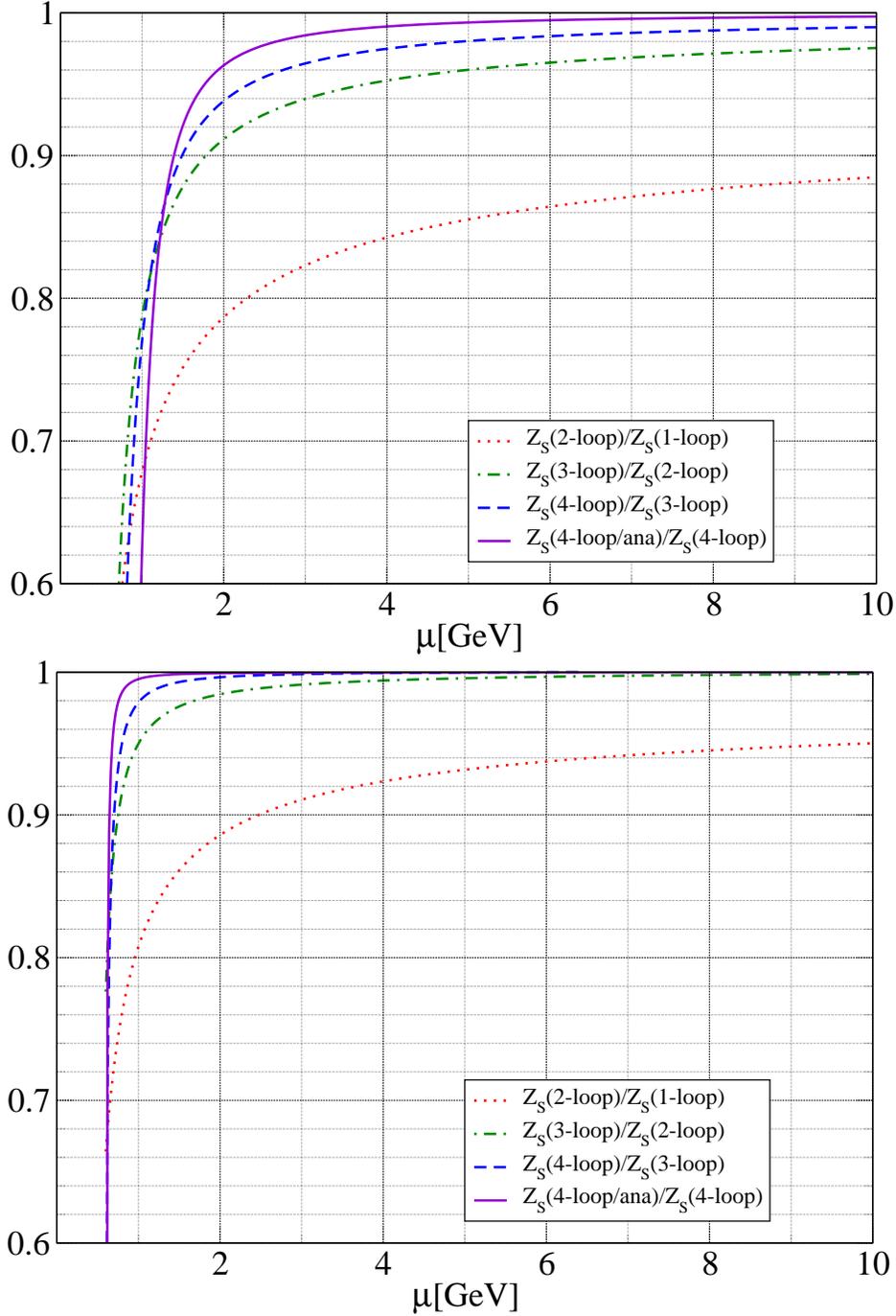

\centering
\vspace{-2mm}
\includegraphics[width=12.3cm]{physpoint_prd.figs/rirun.eps}\\
\includegraphics[width=12.3cm]{physpoint_prd.figs/msbrun.eps}
\vspace{-2mm}
\caption{\label{fig:ch_figs}\sl
Ratio of the perturbatively evaluated $Z_S^\mr{RI}(\mu)$ (top panel) and
$Z_S^{\MSbar}(\mu)$ (bottom panel) at different loop orders. The
renormalization group equations have been numerically integrated, using 1-loop
through 4-loop anomalous dimensions. To estimate the remaining uncertainty in
the 4-loop running, we employ the analytic expression at 4-loop level
\cite{Chetyrkin:1999pq}, which differs from the numerically integrated one by
5-loop effects. In the labels this is called ``4-loop/ana''.}
\end{figure}


In order to obtain tree-level $O(a)$-improved results with Wilson fermions, one
has to improve not only the action, but also the interpolating fields.
For standard correlators this has been discussed in the previous two
subsections.
In addition, in the RI/MOM procedure, one has to remove an $O(a)$ contact term
in the quark propagator \cite{Martinelli:1994ty}.
We apply here the trace subtraction described in
\cite{Becirevic:1999kb,Capitani:2000xi,Martinelli:2001ak,Maillart:2008pv},
which has the added benefit of greatly improving the signal to noise ratio.
This subtraction is implemented by replacing the condition (\ref{RI_condition})
by one in which the modified propagator
$\bar{S}(p)\!=\!S(p)\!-\!\mr{Tr}(S(p))/4$ is used to define the amputated
Green's function, where the trace is in spinor space.

In order to reliably extract the renormalization constants and to convert
the resulting quark masses $m^\mr{RI}(\mu)$ to other schemes without loosing
precision, several conditions should be met:
\begin{itemize}
\vspace*{-2mm}
\itemsep-1mm
\item[($a$)]
the scale $\mu$ at which we take the continuum limit of the RI/MOM renormalized
masses needs to be substantially below the momentum cutoff of the coarsest
lattice $\mu\!\ll\!2\pi/a$,
\item[($b$)]
the conversion to a perturbative scheme has to be done at a scale $\mu'$ which
is sufficiently large, such that perturbation theory is reliable, i.e.\ at
$\mu'\!\gg\!\Lambda_\mr{QCD}$,
\item[($c$)]
the effect of the chiral extrapolation $m\!\to\!0$ needs to be fully controlled.
\vspace*{-2mm}
\end{itemize}
The difficulty to fulfill, in one simulation, the first two conditions is
sometimes referred to as the ``window problem'' of the RI/MOM procedure.
In the following we show how we can simultaneously satisfy all three
requirements.

Ad ($a$):
To renormalize our quark masses and to extrapolate them to the continuum we
choose a convenient renormalization scale $\mu\!=\!2.1\GeV$.
This scale satisfies $\mu\!<\!\pi/(2a)$ for all our lattices (on the coarsest
one this figure is about $2.7\GeV$).
When plotting the running of $Z_{S,\be}(\mu)$ at different $\be$ on top of
each other (see Fig.\,\ref{fig:ch_figr}), one finds that discretization effects
are below our statistical accuracy in this region, and that the form of the
running is almost identical for our five $\be$ values.

Ad ($b$):
With the procedure described above and by taking the continuum limit we obtain
a fully nonperturbative determination of the quark masses in the RI/MOM scheme
at $\mu\!=\!2.1\GeV$.
In principle, we could stop here, quoting this as our main result.
However, if one wants to convert this result to another scale or another
scheme, it is evident from Fig.\,\ref{fig:ch_figs} that doing so perturbatively
would introduce an uncertainty in the $1\!-\!2\%$ range.
Therefore we use our renormalization data to run our quark mass results,
nonperturbatively, to the scale $\mu'\!=\!4\GeV$, where this perturbative
uncertainty is in the $0.5\%$ range and hence subdominant.
At $\mu'\!=\!4\GeV$ we still have 3 different $\be$ values which satisfy the
condition $\mu'\!<\!\pi/(2a)$.
More specifically, we use our data to extrapolate the ratio
$Z_{S,\be}^\mr{RI}(\mu)/Z_{S,\be}^\mr{RI}(\mu')$ to the continuum, with an
extremely mild effect (as one can see from Fig.\,\ref{fig:ch_figr}, in this
interval the three curves lie essentially on top of each other), and the
resulting ratio provides us with the nonperturbative running of the scalar
renormalization constant, in the continuum, between $\mu\!=\!2.1\GeV$ and
$\mu'\!=\!4\GeV$.
Accordingly
\beq
R_S^\mr{RI}(\mu,4\GeV)=\lim_{\be\to\infty}
\frac{Z_{S,\be}^\mr{RI}(4\GeV)}{Z_{S,\be}^\mr{RI}(\mu)}
\label{def_RsubS}
\eeq
is the continuum extrapolated ratio which allows us to evolve data from
\emph{all} our lattices, including the coarser ones, to $\mu'\!=\!4\GeV$, where
we perform the final continuum extrapolation.
Through this procedure we obtain fully nonperturbatively renormalized quark
masses in the RI/MOM scheme at $\mu'\!=\!4\GeV$, which represent our main
result.
For the reader's convenience we also convert them to other schemes.
To this end we use 4-loop perturbative running to convert to the RGI framework
(where we use the conventions of \cite{Garden:1999fg} with $b_0,d_0$ adjusted
to $N_f\!=\!3$), and subsequently to the $\MSbar$ scheme (which is
perturbatively defined).

Ad ($c$):
As mentioned above, the RI/MOM scheme is a massless renormalization scheme.
Since the dedicated $N_f\!=\!3$ simulations as listed in
Tab.\,\ref{tab:runs_nf3} use finite quark masses (roughly in the range
$m_s^\mr{phys}/3\!<\!m\!<\!m_s^\mr{phys}$), we have to perform a chiral
extrapolation at some point.
In the procedure described in the previous paragraph, the numerical data
for $Z_{S,\be,m}^\mr{RI}(\mu)$ were first extrapolated to the chiral limit to
give $Z_{S,\be}^\mr{RI}(\mu)$.
Based on this the renormalized quark masses $m_\be^\mr{RI}(\mu)$ and the
ratios $Z_{S,\be}(\mu')/Z_{S,\be}(\mu)$ were extrapolated to the continuum, as
detailed in (a) and (b), respectively.
To give a reliable estimate of the systematic uncertainties involved, we
supplement this procedure with a second one where we interchange the order of
limits.
Technically, this means that one defines an intermediate MOM scheme, which is
not a massless one, but instead based on a fixed reference quark mass.
We use $m_\mr{ref}^\mr{RGI}\!=\!70\MeV$, since, for all $\be$, this value can
be reached by interpolation.
In this scheme the renormalized light and strange quark masses are determined
at the scales $\mu\!\in\!\{1.3,2.1\}\GeV$, and extrapolated to the continuum.
This yields $m_{ud,m_\mr{ref}}^\mr{MOM}(\mu)$ and ditto for $m_s$.
Staying in this massive scheme, these quark masses are evolved to the scale
$\mu'\!=\!4\GeV$.
In this step a fully controlled continuum extrapolation can be performed, since
we have three lattice spacings satisfying $\mu'\!<\!\pi/(2a)$.
At this point we have the renormalized quark mass in the form
\beq
m_{ud,m_\mr{ref}}^\mr{MOM}(\mu')=
m_{ud,m_\mr{ref}}^\mr{MOM}(\mu)\cdot
\frac{Z_{S,m_\mr{ref}}(\mu)}{Z_{S,m_\mr{ref}}(\mu')}
\label{result_MOM}
\eeq
where either factor has been extrapolated to the continuum.
In the last step, we switch from the intermediate massive MOM scheme to the
massless RI/MOM scheme by multiplying (\ref{result_MOM}) with the continuum
extrapolated ratio $Z_{S,m_\mr{ref}}(\mu')/Z_S(\mu')$.
This yields the same $m_{ud}^\mr{RI}(\mu'),m_s^\mr{RI}(\mu')$ as before, except
that the order of limits has been interchanged.
Note that all continuum extrapolations are entirely flat and the effect of the
mass extrapolation is about 1\%, implying that all limiting procedures are
fully controlled.
Having obtained our main result, the RI/MOM masses at $\mu'\!=\!4\GeV$, we
can transform them to other schemes as described under ($b$).


\subsection{Summary of RI/MOM renormalization}

Let us summarize this section.
We compute the quark masses $m_{ud}^\mr{phys}$ and $m_s^\mr{phys}$ through the
``ratio-difference method'' in the RI/MOM scheme at the scale $\mu'\!=\!4\GeV$,
nonperturbatively and with extrapolation to the continuum.

The mild quark mass dependence of the renormalization factors is eliminated
through a chiral extrapolation.
Also cut-off effects are removed through a continuum extrapolation.
In this step we are extremely conservative -- we do not only consider the
formally leading cut-off effects $O(\alpha a)$, but also subleading effects
proportional to $O(a^2)$, counting the spread towards the final systematic
error (see Sec.\,14).
We think this is necessary, since even with a set of $5$ lattice spacings, we
cannot exclude the possibility that the subleading $O(a^2)$ cut-off effects
largely affect the continuum extrapolation.
If we were to consider only the leading $O(\alpha a)$ cut-off effects, our
systematic error would be significantly smaller.

The quark masses in the RI/MOM scheme at the scale $\mu'\!=\!4\GeV$ are our
main result, obtained in a way which guarantees that they are truly
nonperturbative.
Using perturbation theory in a regime where it is well behaved, we convert them
to the universal RGI prescription and subsequently to the perturbatively
defined $\MSbar$ scheme at the scale $\mu\!=\!2\GeV$.


\section{Quenched overall check}


To demonstrate that our 2-step HEX smeared clover action (\ref{def_action})
and the nonperturbative renormalization of the quark mass yield reliable
results in the continuum, we repeat the quenched benchmark calculation
\cite{Garden:1999fg} of the quantity $m_s\!+\!m_{ud}$, using our setup.

\begin{table}[!tb]
\centering
\begin{tabular}{|cc|c|}
\hline
$\be$ & $L^3\!\times\!T$ & $(m_s\!+\!m_{ud})r_0$ \\
\hline
5.7366 & $12^3\!\times\!24$ & 0.3070(50) \\
5.8726 & $16^3\!\times\!32$ & 0.2801(50) \\
5.9956 & $20^3\!\times\!40$ & 0.2758(52) \\
6.1068 & $24^3\!\times\!48$ & 0.2654(42) \\
6.3000 & $32^3\!\times\!64$ & 0.2685(29) \\
\hline
\end{tabular}
\caption{\label{tab:quenched}\sl
Details of the quenched overall test. The quark masses are in the $\MSbar$
scheme at $2\GeV$.}
\end{table}

\begin{figure}[!p]
\centering
\includegraphics[width=13cm]{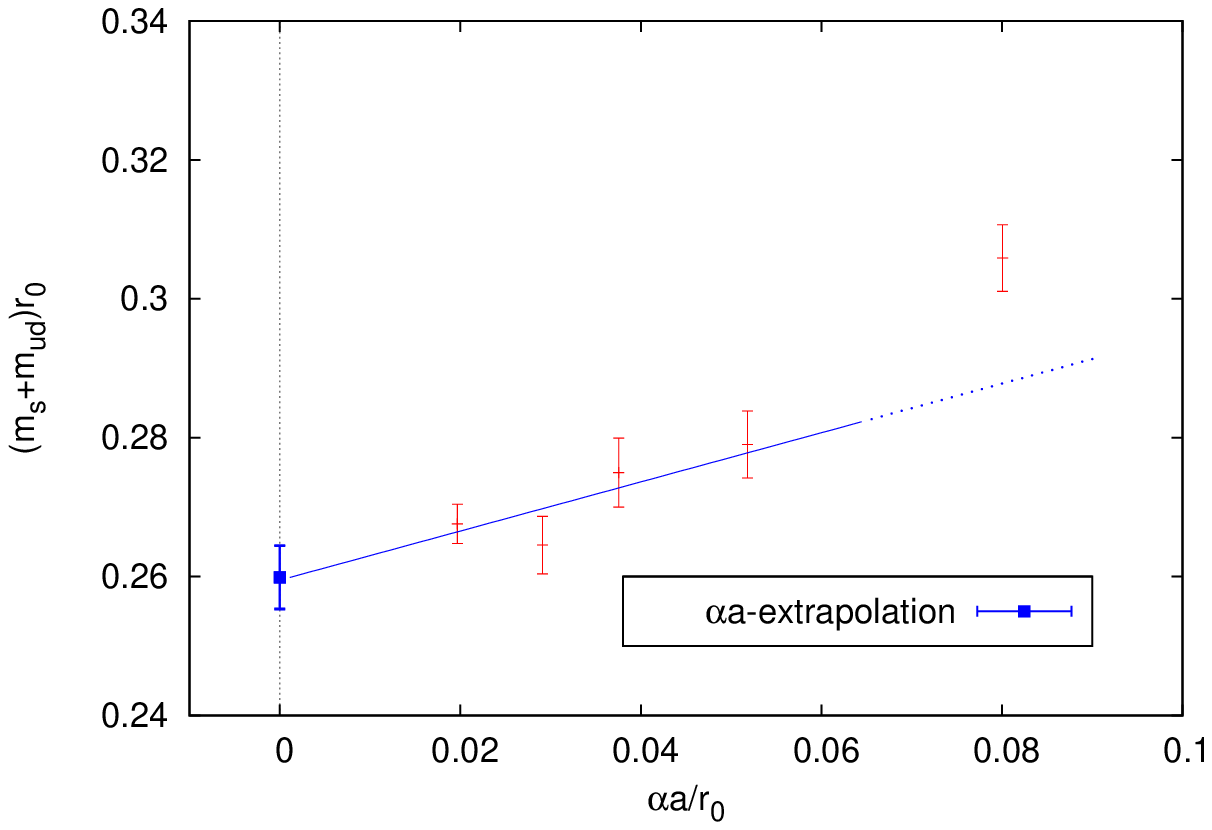}
\includegraphics[width=13cm]{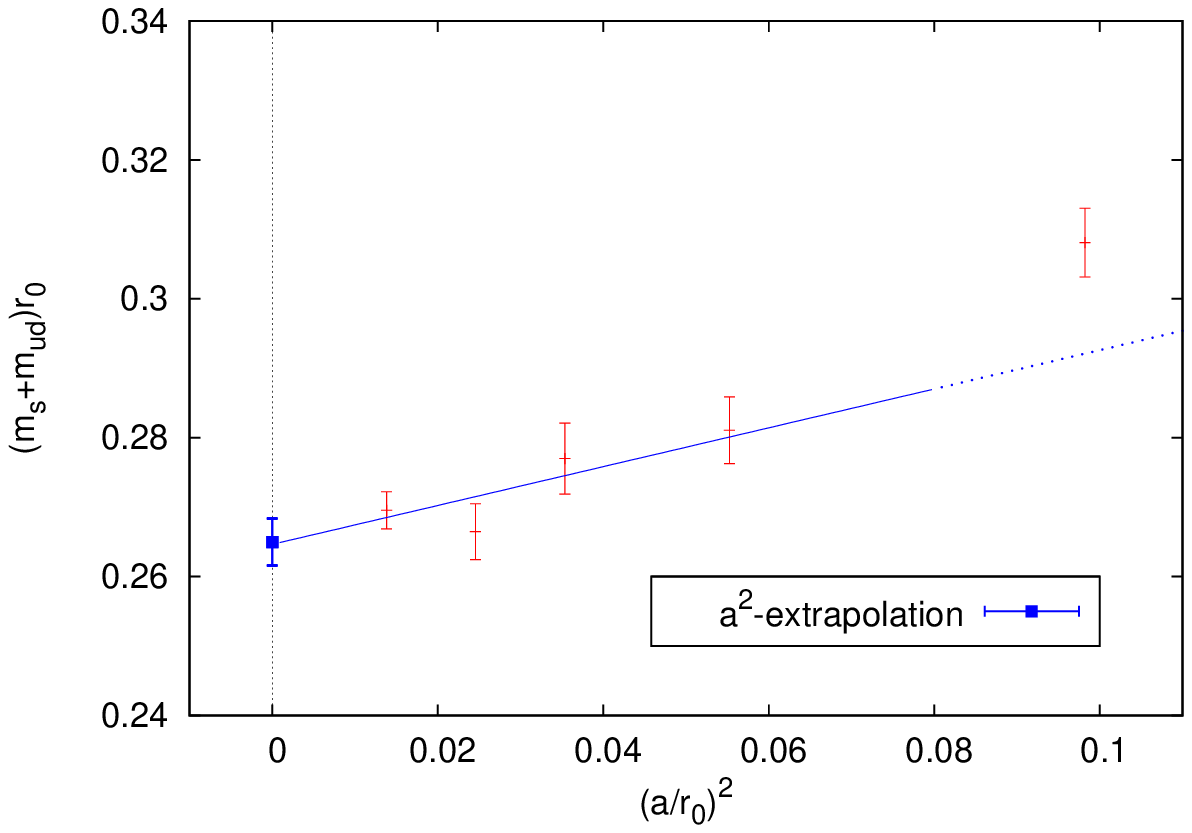}
\vspace{-2mm}
\caption{\label{fig:quenched}\sl
Quenched continuum extrapolation of $(m_s\!+\!m_{ud})r_0$ in the $\MSbar$
scheme at $\mu\!=\!2\GeV$, assuming $O(\alpha a)$ [top] or $O(a^2)$ [bottom]
scaling. One data-point outside the scaling regime ($\be\!=\!5.7366$) is shown.
The difference counts towards the systematic error (see text for details).}
\end{figure}

We use pure Wilson glue at five couplings between $\be\!=5.7366$ and
$\be\!=\!6.3$, each time saving about $600$ well-decorrelated configurations
for the analysis (i.e.\ 200 for the $Z$-factors and 400 for the masses).
The couplings and geometries have been chosen to realize a fixed physical box
size of about $L\!=\!1.84\fm$, see Tab.\,\ref{tab:quenched} for details.
On each set at least 4 quark masses are used to safely interpolate to the point
$M_P\,r_0\!=\!1.229$, where $M_P$ is the pseudoscalar meson mass and where the
numerical value has been chosen to match $\Mka^\mr{phys}r_0$ with
$r_0\!=\!0.49\fm$ \cite{Sommer:1993ce}.

The computation closely follows the dynamical case.
We renormalize the VWI quark mass sum with the methods described in the
previous section, and we use the same procedure to convert to the $\MSbar$
scheme.
In more detail, we begin with measuring $m^\mr{PCAC}$ as a function of the bare
mass $m^\mr{bare}$, which shows a linear relationship.
The intercept with the $x$-axis yields $m^\mr{crit}$ and thus $m^\mr{W}$
as defined in (\ref{def_vwi}).
Next, we determine $Z_S(\mu)$ via RI/MOM \cite{Martinelli:1994ty} to obtain
$m^\mr{VWI}$ according to (\ref{def_vwi}).
It is easy to see that this is the flavor non-singlet $Z_S(\mu)$, since all
quark disconnected contributions vanish in the quenched theory.
In close analogy with our phenomenological analysis, we choose the matching
scales $\mu\!=\!2.1\GeV$ and $\mu'\!=\!3.5\GeV$.
Combining the continuum extrapolated ratio $Z_S(\mu')/Z_S(\mu)$ with
$Z_{S,\be}(\mu)$, we obtain $Z_{S,\be}(\mu')$ and the renormalized mass
$m^\mr{VWI}(\mu')$ in the RI/MOM scheme at the scale $\mu'\!=\!3.5\GeV$.
Finally, we use perturbation theory to convert to the $\MSbar$ scheme at
$2\GeV$ scale.
The result is identified with $m_s\!+\!m_{ud}$ in this scheme, at the given
lattice spacing, and multiplied with $r_0$ to obtain a dimensionless quantity
(cf.\,Tab.\,\ref{tab:quenched}).
We find that we can extrapolate these values linearly in $\alpha a$ or $a^2$,
with the data showing a slight preference for the latter option, as can be
seen from the two panels in Fig.\,\ref{fig:quenched}.
Using the machinery for propagating both statistical and systematic errors
that will be described in Sec.\,14, the combined result in the continuum reads
$(m_s\!+m_{ud})r_0=0.2609(39)(28)$ in the $\MSbar$ scheme at $\mu\!=\!2\GeV$.





Our result is in perfect agreement with the continuum value
$(m_s\!+\!m_{ud})r_0=0.261(9)$ quoted by the ALPHA collaboration
\cite{Garden:1999fg}.
It is consistent, within less than $1\si$, with the result $0.274(18)$ given by
JLQCD \cite{Aoki:1999mr} and, within less than $2\si$, with the value
$0.312(28)$ obtained in a computation with quenched overlap fermions that
includes a continuum extrapolation \cite{Durr:2005ik}.
There is some tension with the result $0.293(6)$ by CP-PACS \cite{Aoki:2002fd},
but one should keep in mind that the systematic uncertainty due to the
perturbative renormalization is not included in their error.
In short, we find good agreement with the most precise results in the
literature.
We take this as evidence that the renormalization procedure described in
Sec.\,11 yields reliable results.


\section{Using dispersive input to obtain $m_u$ and $m_d$}


For decades the most reliable source of information on light quark masses
has been current algebra, in particular in its modern form, known as
Chiral Perturbation Theory (ChPT).
A major drawback of this framework is that only information on quark mass
ratios can be extracted, not on absolute values.
This is a consequence of the fact that in the chiral Lagrangian all quark
masses appear in the combination $B_0m_q$ and the condensate parameter $B_0$
does not occur in any other instance.
We have determined $m_{ud}\!\equiv\!(m_u\!+\!m_d)/2$ and $m_s$ in MeV units.
Accordingly, by comparing our value of the ratio $m_s/m_{ud}$ to theirs, we can
learn something about the convergence pattern of SU(3) ChPT.
Furthermore, one may combine our values for $m_{ud}$ and $m_s$ with the best
available information on another ratio ($Q$, see below) to obtain a result for
the individual quark masses $m_u, m_d$.


\subsection{Comparing our value of $m_s/m_{ud}$ to the one in ChPT}

As a starting point one might ignore higher-order terms in the chiral expansion
and electromagnetic corrections all together.
Upon identifying the left-hand sides in
\bea
M_\pi^2&=&B_0(m_u+m_d)
\\
M_{K^\pm}^2&=&B_0(m_u+m_s)
\\
M_{K^0}^2&=&B_0(m_d+m_s)
\\
M_\eta^2&=&B_0(m_u+m_d+4m_s)/3
\eea
with the experimentally measured meson masses%
\footnote{Pseudoscalars without superscript refer to isospin averages:
$M_\pi^2\!=\!\frac{1}{2}(M_{\pi^\pm}^2\!+\!M_{\pi^0}^2)$,
$M_K^2\!=\!\frac{1}{2}(M_{K^\pm}^2\!+\!M_{K^0}^2)$.},
one obtains three predictions.
On the one hand, the Gell-Mann--Okubo relation
\beq
3\Met^2+\Mpi^2\simeq2M_{K^\pm}^2+2M_{K^0}^2
\eeq
evaluates to $0.919\GeV^2\simeq0.983\GeV^2$, which amounts to a 7\% accuracy.
On the other hand
\bea
(M_{K^\pm}^2+M_{K^0}^2)/\Mpi^2&=&(m_s+m_{ud})/(m_{ud})
\label{quarkmassratio1}
\\
\Met^2/\Mpi^2&=&(2m_s+m_{ud})/(3m_{ud})
\label{quarkmassratio2}
\eea
yield $m_s/m_{ud}\simeq25.1$ and $m_s/m_{ud}\simeq23.4$, respectively.
This spread suggests again a precision of a few percent.
Upon noticing that the $\eta$ undergoes significant mixing with the $\eta'$
and, as a result, that (\ref{quarkmassratio1}) should be preferred over
(\ref{quarkmassratio2}), one arrives at the estimates
\bea
{m_u\ovr m_d}&=&
{M_{K^\pm}^2-M_{K^0}^2+\Mpi^2 \ovr M_{K^0}^2-M_{K^\pm}^2+\Mpi^2}\simeq0.66
\\
{m_s\ovr m_d}&=&
{M_{K^\pm}^2+M_{K^0}^2-\Mpi^2 \ovr M_{K^0}^2-M_{K^\pm}^2+\Mpi^2}\simeq20.8
\eea
which do not take into account electromagnetic contributions to isospin
breaking.


The chiral framework may be extended to include interactions with photons.
At leading order in $\alpha_\mr{em}$ and in the $3$-flavor chiral limit the
electromagnetic contribution to the excess of the charged kaon mass squared is
the same as for the pion, i.e.\
$[M_{\pi^\pm}^2-M_{\pi^0}^2]_\mr{em}=[M_{K^\pm}^2-M_{K^0}^2]_\mr{em}$,
known as ``Dashen's theorem'' \cite{Dashen:1969eg}.
This leads to the improved relations%
\footnote{The numerical values are based on the latest edition of the PDG
\cite{Nakamura:2010zzi} and differ from those given in \cite{Leutwyler:1996qg}.}
\cite{Leutwyler:1996qg}
\bea
{m_u\ovr m_d}&=&
{M_{K^\pm}^2-M_{K^0}^2+2M_{\pi^0}^2-M_{\pi^\pm}^2 \ovr
 M_{K^0}^2-M_{K^\pm}^2+M_{\pi^\pm}^2}\simeq0.56
\\
{m_s\ovr m_d}&=&
{M_{K^\pm}^2+M_{K^0}^2-M_{\pi^\pm}^2 \ovr
 M_{K^0}^2-M_{K^\pm}^2+M_{\pi^\pm}^2}\simeq20.2
\eea
which account for electromagnetism at leading order (LO) in the chiral
expansion.
From this one obtains
$m_s/m_{ud}=2/(m_d/m_s+m_u/m_d\!\cdot\!m_d/m_s)\simeq25.9$
as the LO result in ChPT.
Comparing this to our value (\ref{result_ratio}) [see below] indicates that
--~for this quantity~-- subleading contributions yield only about 6\% of the
total result.


\subsection{Using dispersive information on $Q$ to split $m_{ud}$ into $m_u$
and $m_d$}


As mentioned in the previous subsection, ChPT is well suited to address the
ratios $m_s/m_d$ and $m_u/m_d$.
A way to encode such information on quark mass ratios which, from the ChPT
viewpoint, is particularly robust is to introduce the double ratio
\beq
Q^2\equiv{m_s^2-m_{ud}^2 \ovr m_d^2-m_u^2}
\label{def_Q}
\eeq
since this quantity is unaffected by next-to-leading order (NLO) effects in
the chiral expansion.
Modulo a tiny correction, (\ref{def_Q}) can be put into a form known as
``Leutwyler's ellipse'' \cite{Leutwyler:1996qg}
\beq
{1\over Q^2}\Big({m_s\over m_d}\Big)^2+\Big({m_u\over m_d}\Big)^2=1
\label{leutwyler}
\eeq
and relying on Dashen's theorem \cite{Dashen:1969eg} or refinements thereof
(see e.g.\ \cite{FLAG}), one might attempt to determine the value of $Q$ from
the masses of the charged and neutral kaon and pion.


Since we intend to use (\ref{def_Q}) to predict the isospin splittings in QCD
(i.e.\ without electromagnetism), it seems more advisable to build on the long
tradition in the phenomenological literature to determine $Q$ from the rate for
$\eta\to3\pi$ decays or from the branching ratio of $\psi'\to\psi\pi^0$ versus
$\psi'\to\psi\eta$ decays.
The former amplitude seems particularly interesting, as it violates isospin,
while being barely affected by electromagnetic corrections
\cite{Ditsche:2008cq}.
Evidently, this renders it sensitive to the effect of $m_d\!-\!m_u\!\neq\!0$,
which is exactly what we are interested in.
In the following, we restrict ourselves to the dispersive treatment of the
$\eta\to3\pi$ amplitude, as given by Kambor-Wiesendanger-Wyler
\cite{Kambor:1995yc}, Anisovich-Leutwyler \cite{Anisovich:1996tx}, and
Colangelo-Lanz-Passemar \cite{Colangelo:2009db}.
In the first place we note that the central value found in these works has been
remarkably consistent over one and a half decades.
Let us also emphasize that a dispersive treatment is, conceptually, as much
from first principles as a lattice computation~-- dispersion theory rests
exclusively on the axioms of quantum field theory.
In a world with perfect experimental data, this would be the complete story.
However, with presently available data, additional input is required
(see e.g.\ \cite{Colangelo:2009db}).
To account for such provisional effects, Leutwyler has assigned his estimate
$Q\!=\!22.3(8)$ \cite{Leutwyler:2009jg} a much larger error bar than claimed
in some of the publications it is based on.
In our view this is the most accurate value available, if one is not willing to
resort to model calculations, and we shall thus stay content with its rather
conservative error bar.

We now extend our lattice determinations of $m_{ud}$ and $m_s/m_{ud}$ to all
three quark masses, using this dispersive information.
Upon rewriting (\ref{def_Q}, \ref{leutwyler}) in the form
\beq
{1\ovr Q^2}=4\;\Big({m_{ud}\ovr m_s}\Big)^2\;{m_d-m_u\ovr m_d+m_u}
\label{rewritten}
\eeq
it follows that the above-mentioned value of $Q$ and our lattice result
\beq
\frac{m_s}{m_{ud}}=27.53(20)(08)
\label{result_ratio}
\eeq
yield the light quark mass asymmetry parameter
\beq
\frac{m_d-m_u}{m_d+m_u}=0.381(05)(27)
\label{def_asy}
\eeq
where the error on $Q$ is considered a systematic error.
As an aside we mention that this asymmetry parameter is equivalent to
$m_u/m_d=0.448(06)(29)$.
Combining (\ref{def_asy}) with our result $m_{ud}=3.503(48)(49)\MeV$, we obtain
\beq
m_u=2.17(04)(10)\MeV,\quad
m_d=4.84(07)(12)\MeV
\eeq
with all masses given in the RI/MOM scheme at the scale $\mu\!=\!4\GeV$.
These values and our original results for $m_s$ and $m_{ud}$ (along with their
counterparts in the RGI and $\MSbar$ schemes) are summarized in
Tab.\,\ref{tab:mass_results} (see Sec.\,15) and quoted in \cite{Durr:2010??}.

To summarize the technical part, let us say that we have determined $m_u$ and
$m_d$, based on our lattice value of $m_{ud}$, our lattice value of the
ratio $m_s/m_{ud}$ and the dispersive treatment of $Q$.
Given that our simulation points bridge the physical values of $m_{ud}$ and
$m_s$ (cf.\ Sec.\,5), the chiral framework is no longer needed in the first two
quantities, and the use of ChPT is thus limited to a subdominant contribution
in a mostly dispersive framework to determine $Q$.


\subsection{Physics implication, robustness issues and precision outlook}

Physicswise, an important conclusion is that our result (\ref{def_asy}) for the
light quark mass asymmetry parameter excludes a vanishing up-quark mass by
$22.1$ standard deviations.
This is a consequence of the dispersive determination of $Q$ being entirely
inconsistent with $13.8$, the value of $Q$ which relation (\ref{rewritten}) and
our result for $m_s/m_{ud}$ would enforce if $m_u\!=\!0$.
As can be seen from (\ref{rewritten}), the asymmetry parameter depends strongly
on the ratio $m_s/m_{ud}$, which is the quantity that we have determined to
sub-percent precision.
The bottom line is that our precise lattice results and the dispersive
processing of phenomenological information which excludes very large
corrections to Dashen's theorem, when combined, rule out the simplest proposed
solution to the so-called ``strong CP problem''.
This corroborates previous findings \cite{Leutwyler:1996qg}.

Note that the way in which we have used phenomenological information is
designed to make sure that the so-called ``Kaplan-Manohar ambiguity'' is
circumvented in our derivation of $m_u$ and $m_d$.
This ambiguity expresses the fact that a redefinition of the quark condensate
and of certain low-energy constants allows one to move on Leutwyler's ellipse
\cite{Kaplan:1986ru}.
It represents an accidental symmetry of those Green's functions in the
effective theory which determine pseudoscalar masses, scattering amplitudes and
matrix elements of the vector and axial-vector currents
\cite{Leutwyler:1996qg}.
However, the \emph{aspect ratio} of Leutwyler's ellipse is not affected by this
ambiguity, and it is this shape information%
\footnote{We remark that $Q$ as defined in (\ref{def_Q}) picks up, under a
Kaplan-Manohar transformation, terms of order NNLO and a change proportional
to $m_d\!-\!m_u$. The latter ``deficiency'' could be cured by defining
$Q_1^2\!=\!(m_s^2\!-\!m_d^2)/(m_d^2\!-\!m_u^2)$ \cite{Leutwyler_private}.
Note, however, that the numerical difference between $Q$ and $Q_1$ [or $Q_2$,
the quantity that shows up in (\ref{rewritten})] is about one permil, i.e.\
more than an order of magnitude smaller than the uncertainty that we have
assigned to $Q$.}
which is encoded in $Q$.
In consequence, relation (\ref{rewritten}) ensures that the high precision that
we have reached in $m_s/m_{ud}$, together with the robust value of $Q$ that we
use, leads to a determination of the asymmetry parameter (\ref{def_asy}) and
thus of the individual $u$ and $d$ quark masses which is unaffected by the
Kaplan-Manohar ambiguity.

We stress that, in our view, there is not much conceptual difference between
using only $\Mpi,\Mka$ as input quantities versus including $Q$, too.
To compute $m_{ud},m_s$, we needed two (polished) experimental input numbers
to adjust the average light and the strange quark masses, apart from $M_\Omega$
to set the overall scale (cf.\ Sec.\,4).
To compute $m_u,m_d,m_s$, evidently, we need a third one, and we are well
advised to choose one which is sensitive to the effect we want to quantify.
We select $Q$ for its large sensitivity to QCD-induced isospin breaking, thus
requiring very little theoretical polishing, and for this little bit resting on
dispersion theory which is well founded.
Still, there is room for improvement, as can be seen from the fact that our
value of $m_{ud}$ had 2\% precision, while $m_u$ and $m_d$ have only 5\% and
3\% accuracy, respectively.
The problem is that the current value of $Q$ determines the asymmetry parameter
(\ref{def_asy}) to only about 7\% precision.
While improvements on the value of $Q$ obtained in this way may be possible
\cite{Colangelo:2009db}, reaching accuracies of $m_u,m_d$ below the few percent
level will most probably require a different approach, even more heavily based
on lattice field theory.
Indeed, once simulations become available with $N_f\!=\!1\!+\!1\!+\!1$ physical
quark flavors (i.e.\ with non-degenerate up, down, and strange quark masses,
each of which is taken at its physical value) and with an additional abelian
gauge field%
\footnote{For recent progress in this field see e.g.\ \cite{Blum:2010ym}.}
to account for electromagnetic interactions, it will become possible to take
full advantage of the very accurately known $K^+$ and $K^0$ masses to determine
$m_u$ and $m_d$ with even higher precision.


\section{Assessment of systematic errors}


Our approach is to establish one global fit to interpolate our $11+12+9+9+6=47$
simulations at $5$ different lattice spacings (cf.\ Tab.\,\ref{tab:params}) to
the physical mass point (i.e.\ physical $\Mpi$ and $\Mka$) and to
extrapolate to zero lattice spacing (i.e.\ $a\!\to\!0$).
In order to obtain a reliable estimate of the systematic error involved, we
repeat the entire analysis with a large selection of interpolation formulae,
mass cuts, discretization terms, fit ranges, and renormalization procedures.



In order to extrapolate or interpolate a given quantity to the physical quark
mass point, one needs to expand it around some pion and kaon mass point.
Often the $N_f\!=\!2$ or $N_f\!=\!3$ chiral limit is chosen as an expansion
point and hence SU(2) or SU(3) ChPT \cite{Gasser:1983yg,Gasser:1984gg} as the
theoretical framework.
Expressing the dependence on the light quark mass as a dependence on $\Mpi$,
this kind of expansion leads, for a quantity which vanishes in the chiral
limit, to a quadratic term $\propto\!\Mpi^2$ and higher order chiral logs,
e.g.\ $\propto\!\Mpi^4\log(\Mpi^2/\Lambda^2)$, with known prefactors but
unknown scale $\Lambda$.
In many cases the practical usefulness of knowing the prefactors is limited,
since they contain other quantities (e.g.\ the axial coupling $g_A$ for octet
baryons) which may not be available from the same simulation and which one may
not want to borrow from phenomenology.
Furthermore, it is rather difficult for a fit to tell, e.g., a pure $\Mpi^4$
contribution from an $\Mpi^4\log(\Mpi^2/\Lambda^2)$ chiral log.
Accordingly, choosing an expansion point for an interpolation somewhere in the
middle of the region where one has data (or in the middle of the region defined
by the data points and the target point in case of an extrapolation) and using
a simple Taylor expansion in $\Mpi^2$ leads to rather similar results
\cite{Durr:2008zz}.

To flesh out the meaning of these statements, let us consider the quantities of
interest, $m_{ud}$ and $m_s$.
In our analysis we use the NLO mass formulae (\ref{xpt_mpi}) from SU(2) ChPT
\cite{Gasser:1983yg}, albeit in reversed form, so that it expresses $m_{ud}$ as
a function of $\Mpi$.
To the order we are working at, this can be done in several ways [the
difference is an NNLO effect]; we use the relations
\bea
m_{ud}&=&\frac{\Mpi^2}{2B}\cdot\Big\{
1-\frac{1}{2}\frac{\Mpi^2}{(4\pi\Fpi)^2}\log(\frac{\Mpi^2}{\Lambda_3^2})
\Big\}
\cdot\Big(1+c_s\Delta\Big)
\\
m_{ud}&=&\frac{\Mpi^2}{2B}\Big/\Big\{
1+\frac{1}{2}\frac{\Mpi^2}{(4\pi\Fpi)^2}\log(\frac{\Mpi^2}{\Lambda_3^2})
\Big\}
\cdot\Big(1+c_s\Delta\Big)
\eea
where we have introduced a hadronic quantity
\beq
\Delta=2\Mka^2-\Mpi^2-[2\Mka^2-\Mpi^2]^\mr{phys}
\eeq
to parametrize the small deviation of our strange quark mass from its physical
value \cite{Durr:2008zz}.
Alternatively, for the light quark mass we use a Taylor expansion of the form
\beq
m_{ud}=c_1+c_2\Mpi^2+c_3\Mpi^4+c_4\Delta
\eeq
while the strange quark mass is always parametrized as
\beq
m_s=c_5+c_6\Mpi^2+c_7\Delta+c_8\Delta^2
\;.
\eeq
We have tried to augment these formulas by higher order terms, both in $\Mpi^2$
and $\Delta$, but we found those coefficients to be consistent with zero, with
the given precision of our data.
This yields 3 options for the mass interpolation or extrapolation of the
pseudoscalars.
Similarly, for the $\Omega$ baryon that serves to set the scale, a Taylor
ansatz in $\Mpi^2$ and $2\Mka^2\!-\!\Mpi^2$ is used (cf.\ Sec.\,4).
In total we have $3$ functional ansaetze to interpolate our data.

A standard way to test the functional ansatz is to prune the data with mass
cuts.
We use $\Mpi\!<\!\{380,480\}\MeV$ for the scale setting and
$\Mpi\!<\!\{340,380\}\MeV$ for the quark mass determination,
thus a total of 4 mass cuts.

A source of error which, in practice, often proves dominant is the
contamination of the ground state in the two-point correlator by excited
states.
To reduce this contamination we use a Gaussian source and sink with a fixed
width of about $0.32\fm$.
We tested 1-state and 2-state fits, and found complete agreement if the 1-state
fits start at $t_\mr{min}\!\simeq\!0.7\fm$ for the $PP,PA_4,A_4P,A_4A_4$ meson
channels and from $t_\mr{min}\sim0.8\fm$ for the $\Omega$.
In lattice units this amounts to $at_\mr{min}\!=\!\{6,8,9,11,13\}$ for
$\be\!=\!\{3.31,3.5,3.61,3.7,3.8\}$ (and $\sim\!20\%$ later for baryons).
In order to estimate any remaining excited state effects, we repeated
our analysis with an even more conservative meson fit range (starting at
$at_\mr{min}=\{7,9,11,13,15\}$ and again $\sim\!20\%$ later for baryons).
The end of the fit interval was always chosen to be
$at_\mr{max}=2.7\!\times\!at_\mr{min}$ or $T/2-1$ for lattices with a time
extent shorter than $5.4\!\times\!at_\mr{min}$.
In all cases, the fits were performed in a correlated way.
In total this gives 2 different fit ranges to make sure that contamination
by excited states is under control.

As a result of the tree-level value $c_\mr{SW}\!=\!1$ our action has formally
$O(\alpha a)$ cut-off effects.
However, due to the smearing the coefficient in front of this term is small,
and the formally sub-leading $O(a^2)$ contributions might numerically dominate
over the $O(\alpha a)$ part.
To account for this we augment our global fit by cutoff terms which stipulate
either $O(\alpha a)$ or $O(a^2)$ deviation from the continuum.
This ambiguity comes into play in the evolution function (\ref{def_RsubS}) and
in the continuum extrapolation of the quark masses in the RI/MOM scheme, which
yields 4 options.

Besides the variations described above, we consider 3 options in the
nonperturbative renormalization procedure (scale $\mu$, massless versus massive
intermediate scheme), see Sec.\,11.

\begin{table}[!tb]
\centering
\begin{tabular}{|ccc|cccccc|}
\hline
cen.\,val.&$\sigma_\mr{stat}$&$\sigma_\mr{syst}$&
plateau&scale\,set&fit\,form&mass\,cut&renorm.&cont.\\
\hline
  3.503 & 0.048 & 0.049 & 0.330 & 0.034 & 0.030 & 0.157 & 0.080 & 0.926 \\
 96.43  & 1.13  & 1.47  & 0.207 & 0.005 & 0.031 & 0.085 & 0.085 & 0.970 \\
 27.531 & 0.196 & 0.083 & 0.513 & 0.200 & 0.023 & 0.320 &  ---  & 0.771 \\
\hline
\end{tabular}
\normalsize
\caption{\label{tab:errorbudget}\sl
Split-up of the total systematic uncertainty of $m_{ud}^\mr{phys}$,
$m_s^\mr{phys}$ and $m_s^\mr{phys}/m_{ud}^\mr{phys}$ (from top to bottom) into
the various contributions. Entries in columns 1-3 are in MeV and refer to the
RI/MOM scheme at $\mu\!=\!4\GeV$. Columns 4-10 indicate the relative share of
the systematic error given in column 3 (the squares of these numbers add up to
1). The headers of these columns refer to the plateau range in the primary
observables, the overall scale setting, the interpolation ansatz to tune to the
physical mass point, the cut in the pion mass, the details of the
renormalization procedure (read-off scale, chiral extrapolation), and the
continuum extrapolation.}
\end{table}

All of this serves the goal of quantifying potential systematic effects on our
final results.
In addition, there are standard methods to assess the size of the statistical
error.
Apart from the autocorrelation analysis detailed in Sec.\,7, we used different
blocking sizes on our ensembles, ranging from 1 to 10 configurations, where two
adjacent configurations are separated by ten $\tau\!=\!1$ MD updates
(cf.\ Sec.\,5).
Last but not least, we found that artificial thermalization cuts (where we
ignore the first 20-100 configurations of the thermalized ensembles) induce no
noticeable change in our results, and therefore we conclude that possible
residual thermalization effects are irrelevant for the error analysis.

Putting everything together we have
3 ansaetze for the interpolation of the quark masses to the physical point,
4 mass cuts in the scale setting and the quark mass determination,
2 different fit intervals for the primary observables,
4 ansaetze for the continuum extrapolation, and
3 ways of doing the RI/MOM renormalization.
This gives a total of $3\cdot4\cdot2\cdot4\cdot3=288$ analyses.

In order to quote a final result, we follow the procedure used in
\cite{Durr:2008zz}.
It is essential to note that we have no \emph{a priori} reason to favor one of
these fits over another.
Therefore the analysis method should represent the full spread of all
reasonable and theoretically justified treatments of our data.
In other words, we use all $288$ procedures, and weigh the results by the
quality of fit $Q=\Gamma(n/2,\chi^2/2)$ to form a histogram.
Next, we compute the mean and standard deviation of the distribution, and
this yields the central value and the systematic error which we quote.
Finally, we repeat this extensive procedure on 2000 bootstrap samples.
The standard bootstrap error of the mean gives the statistical error.

An additional benefit of our method to treat systematic effects is that we can
temporarily suppress one of the variations considered (i.e.\ abandon one of the
factors who's product leads to the $288$ procedures) to learn about the
contribution of this individual factor to the total error.
The total ``error budget'' compiled in this way is shown in
Tab.\,\ref{tab:errorbudget}.
Evidently, it exhibits the cut-off effects as the dominant source of systematic
uncertainty in our results.

All together, our procedure to assess both statistical and systematic errors is
an extended frequentist method \cite{Nakamura:2010zzi} which was already used
in \cite{Durr:2008zz}.


\section{Summary}


We have carried out a precise determination of the average light quark mass
$m_{ud}\!=\!(m_u\!+\!m_d)/2$ and of the strange quark mass $m_s$, using
nonperturbative $N_f\!=\!2\!+\!1$ lattice QCD and nonperturbative
renormalization throughout.
Our data cover $5$ lattice spacings in the range $0.054\!-\!0.116\fm$, with
pion masses down to $\sim\!120\fm$ and box sizes up to $6\fm$.
This allows for a safe extrapolation to the continuum ($a\!\to\!0$) and to
infinite volume ($L\!\to\!\infty$).

\begin{figure}[!tb]
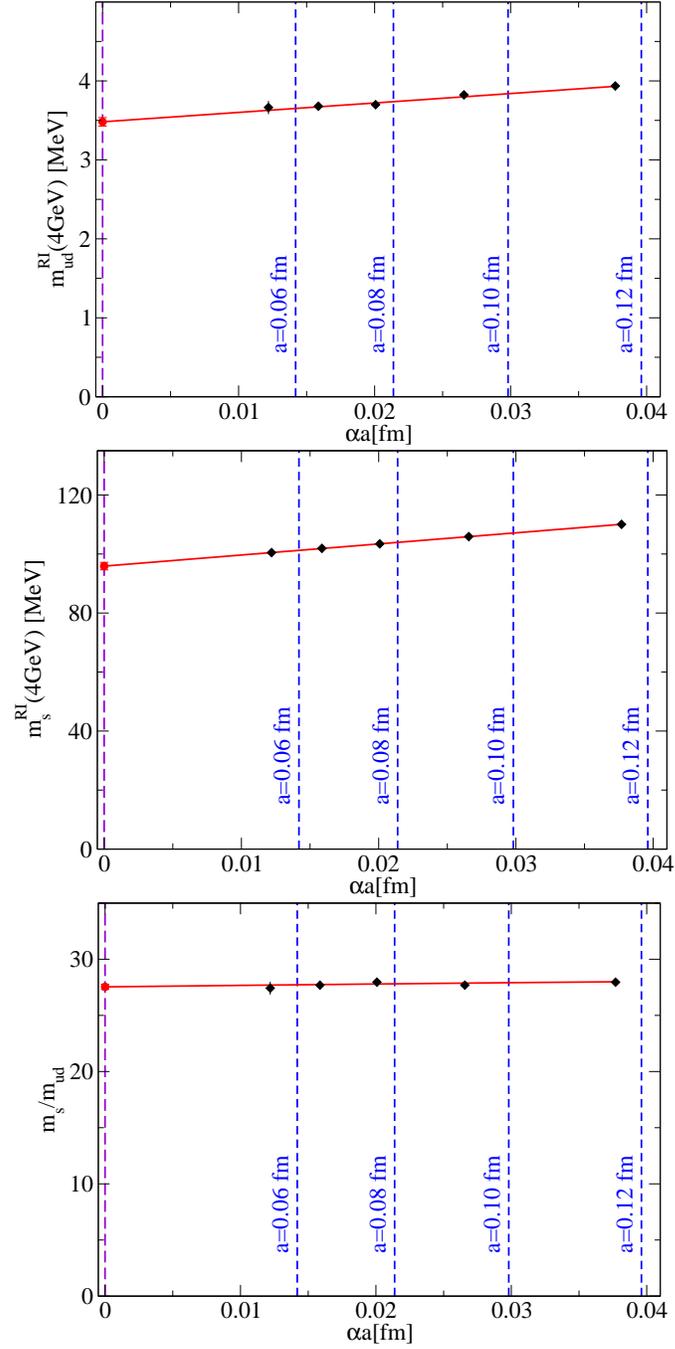

\centering
\includegraphics[width=8.4cm]{physpoint_prd.figs/xml_f2.eps}
\includegraphics[width=8.8cm]{physpoint_prd.figs/xms_f2.eps}\phantom{x}
\includegraphics[width=8.4cm]{physpoint_prd.figs/xmsl_f2.eps}
\caption{\label{fig:mq_final}\sl
Continuum extrapolation of $m_{ud}$ (top), $m_s$ (middle), $m_s/m_{ud}$
(bottom) versus $\alpha a$, for one of our 288 analyses with a good fit
quality (cf.\ discussion in Sec.\,14).}
\end{figure}

We have devised a number of innovative methods, most notably a scheme to
exploit the different renormalization pattern of Wilson and PCAC quark masses
with tree-level $O(a)$-improved clover quarks and a procedure to overcome the
RI/MOM window problem by taking a separate continuum limit of the running of
the scalar density $R_S(\mu,\mu')$.

\begin{table}[!tb]
\centering
\begin{tabular}{|l|cc|cc|}
\hline
  & $m_s$ & $m_{ud}$ & $m_u$ & $m_d$ \\
\hline
$\mbox{RI/MOM}(4\GeV)$ & \,\,96.4(1.1)(1.5) & 3.503(48)(49) & 2.17(04)(10) & 4.84(07)(12) \\
$\mbox{RGI}$           &$\!$127.3(1.5)(1.9) & 4.624(63)(64) & 2.86(05)(13) & 6.39(09)(15) \\
$\MSbar(2\GeV)$        & \,\,95.5(1.1)(1.5) & 3.469(47)(48) & 2.15(03)(10) & 4.79(07)(12) \\
\hline
\end{tabular}
\caption{\label{tab:mass_results}\sl
Renormalized quark masses in the RI/MOM scheme at $\mu\!=\!4\GeV$, and after
conversion to RGI and the $\MSbar$ scheme at $\mu\!=\!2\GeV$. The RI/MOM values
are fully nonperturbative, so the first line is our main result. The first two
columns emerge directly from our lattice calculation, the last two build, in
addition, on dispersive information on $Q$. The precision of $m_s$ and $m_{ud}$
is somewhat below the 2\% level, for $m_u$ and $m_d$ it is about 5\% and 3\%,
respectively. The ratio $m_s/m_{ud}=27.53(20)(08)$ is independent of the scheme
and accurate to better than 1\%.}
\end{table}

Our main result, $m_s$ and $m_{ud}$ in the RI/MOM scheme at renormalization
scale $\mu\!=\!4\GeV$ (cf.\ Tab.\,\ref{tab:mass_results}), is from first
principles and fully nonperturbative.
To ease comparison with the literature, these values are converted to RGI
conventions and, subsequently, to the $\MSbar$ scheme.
In this step reference to perturbation theory is unavoidable, but we do this in
a controlled way, since we show that the 4-loop anomalous dimension of the
scalar density is consistent with our nonperturbative running for
$\mu\gsim4\GeV$.
The ratio $m_s/m_{ud}$ is scheme and scale invariant.
It turns out that our action entails favorable scaling properties not just for
hadron masses, but also for renormalized quark masses, as the plot of a
representative continuum extrapolation in Fig.\,\ref{fig:mq_final} shows.
The combination of using pion masses down to (and even below) the physical
value and having precise and fully nonperturbative renormalization factors
allows us to determine $m_s$ and $m_{ud}$ with a precision of better than 2\%,
and their ratio to better than 1\%.

A determination of the individual light quark masses $m_u$ and $m_d$ by lattice
methods alone is beyond the scope of this paper.
Nevertheless, the precision of our values of $m_{ud}$ and $m_s/m_{ud}$ allows
for a fruitful use of the result of the dispersive analysis of the double ratio
$Q$ (cf.\ discussion in Sec.\,13).
By combining these pieces of information, we obtain values of the individual
quark masses $m_u$ and $m_d$ with a precision of 5\% and 3\%, respectively
(cf.\ Tab.\,\ref{tab:mass_results}).

\bigskip

\noindent
{\bf Acknowledgments}

\smallskip

We used HPC resources from FZ J\"ulich and from GENCI-[IDRIS/CCRT] grant 52275,
as well as clusters at Wuppertal and CPT. This work is supported in part by
EU grants I3HP, FP7/2007-2013/ERC n$^o$ 208740, MRTN-CT-2006-035482
(FLAVIAnet), DFG grant FO 502/2, SFB-TR 55, CNRS GDR 2921 and PICS 4707.


\end{document}